\documentclass[12pt]{article}

\catcode`\@=11
\@addtoreset{equation}{section}

\global\arraycolsep=2pt

\usepackage{mathrsfs,amsbsy,amssymb,latexsym,amsfonts,amsmath,cite}
\usepackage{graphicx,color}

\newcommand{\dd}{{\rm d}}

\newcommand{\gym}{g_{\text{YM}}}
\newcommand{\gs}{g_{\text{s}}}
\newcommand{\pd}{{\cal D} }
\newcommand{\ads}{\text{AdS} }
\newcommand{\tf}{T_{\rm F} }
\newcommand{\diag}{{\rm{diag}} }
\newcommand{\tr}{\text{tr} }
\newcommand{\rc}{r_{\rm c} }

\newcommand{\rt}{r_{\rm t} }
\newcommand{\sym}{{\rm SYM} }
\newcommand{\im}{{\rm Im} }
\newcommand{\ex}{\text{ex}}

\newcommand{\cl}{\text{cl} }
\newcommand{\2}{{\it 2} }

\allowdisplaybreaks

\begin{document}
\begin{flushright}
\parbox{4.2cm}
{KUNS-2552}
\end{flushright}

\vspace*{2cm}

\begin{center}
{\Large \bf A holographic description of the Schwinger effect \\ in a confining gauge theory}
\vspace*{2cm}\\
{\large Daisuke Kawai\footnote{E-mail:~daisuke@gauge.scphys.kyoto-u.ac.jp}, 
Yoshiki Sato\footnote{E-mail:~yoshiki@gauge.scphys.kyoto-u.ac.jp} 
and 
Kentaroh Yoshida\footnote{E-mail:~kyoshida@gauge.scphys.kyoto-u.ac.jp} 
}
\end{center}

\vspace*{1cm}
\begin{center}
{\it Department of Physics, Kyoto University \\ 
Kyoto 606-8502, Japan} 
\end{center}

\vspace{1cm}

\begin{abstract}
This is a review of the recent progress on a holographic description of the Schwinger effect. 
In 2011, 
Semenoff and Zarembo proposed a scenario to study the Schwinger effect 
in the context of the AdS/CFT correspondence. The production rate of quark anti-quark pairs 
was computed in the Coulomb phase. In particular, it provided the critical value of external 
electric field, above which particles are freely created and the vacuum decays catastrophically. 
Then the potential analysis in the holographic approach was invented and 
it enabled us to study the Schwinger effect in the confining phase as well. 
A remarkable feature of the Schwinger effect in the confining phase 
is to exhibit another kind of 
the critical value, below which the pair production 
cannot occur and the vacuum of the system is non-perturbatively stable. The critical value 
is tantamount to the confining string tension. We computed the pair production rate 
numerically and introduced new exponents 
associated with the critical electric fields. 
\end{abstract}

\thispagestyle{empty}
\setcounter{page}{0}
\setcounter{footnote}{0}

\newpage

%
%

\tableofcontents

\section{Introduction}	

The Schwinger effect\cite{Schwinger,HE,Dunne1} 
is a pair creation process in quantum electrodynamics (QED), 
due to an external electric field\footnote{
The Schwinger effect is not intrinsic to QED,  
but it is ubiquitous in quantum field theories coupled with a $U(1)$ gauge field. 
Furthermore, it may be generalized to non-abelian gauge fields like color fields 
in quantum chromodynamics (QCD) \cite{na1,na2,na3}.}.
The production rate (per unit time and volume) is given by
\begin{equation}
\Gamma \sim \exp \left(-\frac{\pi m^2}{eE}\right)\,,
\label{Srate}
\end{equation}
where $m, e $ and $E$ are an electron mass, an elementary electric charge 
and an external electric field, respectively. This formula is computed under 
the weak electric-field ($eE \ll m^2$)\,. 
A typical value of $E$ for which the Schwinger effect becomes significant 
is estimated as 
\begin{equation}
E=E_{\rm Sch}=\frac{m^2}{e} \simeq 1.3 \times 10^{18}\, {\rm V/m}\,.
\end{equation}
Thus $E_{\rm Sch}$ is extremely large in comparison to 
typical values of electric field in table-top experiments. 
For example, a typical value of electric field necessary to ionize an atom is given by 
\begin{equation}
E_{\rm ion}=m^2 e^5 \alpha _{\rm s}^3 \simeq 5.2 \times 10^{11}\, {\rm V/m}\,,
\end{equation}
with the fine-structure constant $\alpha_{\rm s}=e^2/4\pi$\,.
At least so far, in real experiments, the Schwinger effect has not been observed yet, 
but it may be observed in the near future.  
The XFEL project at DESY and the ELI project in Europe plan to 
produce extremely strong electric field very close to $E_{\rm Sch}$. 

\smallskip

The exponential factor in the production rate indicates that the pair production process 
should be described as a non-perturbative phenomenon like a tunneling process in quantum mechanics. 
To make a virtual pair of electron and positron be real particles, it is necessary 
to achieve larger energy than the static energy from an external source. 
A phenomenological potential,   
\begin{equation}
V_{\rm tot}(x)=2m-\frac{\alpha _{\rm s}}{x}-eEx\,,
\end{equation}
leads us to a tunneling picture. 
The potential $V_{\rm tot}(x)$ is composed of three parts: 
1) the static electron mass, 
2) the Coulomb potential with the distance $x$ between electron and positron, and 
3) an energy provided by the external electric field $E$\,. 
For the detail, see e.g.\ chapter 13 in the book\cite{Alvarez}.

\smallskip

\begin{figure}[tbp]
\begin{center}
\includegraphics[scale=.6]{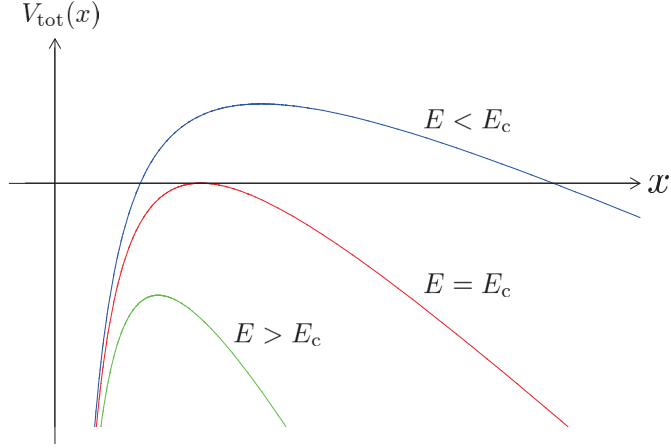}
\end{center}
\vspace*{-0.7cm}
\caption{The potential shapes. For the blue line, the particle production is described as 
a tunneling process. The red line indicates that the potential barrier vanishes 
at a certain value $E=E_{\rm c}$ (critical electric field). 
For the green line, the vacuum is catastrophically unstable.} 
\label{tunnel:fig}
\end{figure}

The potential shapes are depicted in Fig.\ \ref{tunnel:fig}\,. 
When $E$ is not so large, one can see the potential barrier (the blue line in Fig. \ref{tunnel:fig}). 
But the particle can penetrate it quantum mechanically, the particle creation can 
be described as a tunneling phenomenon. This is nothing but the Schwinger effect. 
It would be worth estimating the production rate roughly
in the WKB sense. The triangle approximation of the potential barrier leads to the 
following exponential factor, 
\[
\Gamma \sim \exp \left( -\frac{m^2}{eE}\right)\,.
\]
This is very close to Schwinger's result (\ref{Srate}), despite the rough estimation.

\smallskip

As $E$ increases, the barrier tends to be lowered and at last it disappears completely 
at a certain value of $E$ (the red line in Fig.\,\ref{tunnel:fig}). 
This value is referred to as the critical electric field $E_{\rm c}$\,.  
The disappearance of the barrier indicates that the QED vacuum becomes unstable 
catastrophically once $E$ has reached $E_{\rm c}$\,. 
It is easy to evaluate this value as 
\[
eE_{\rm c} = \frac{m^2}{\alpha_{\rm s}} \sim 137 m^2\,.
\]
Note that this value is far beyond the weak-field approximation $eE \ll m^2$\,. 
Hence, frankly speaking, it has not been clarified 
whether the catastrophic decay would really occur or not. 

\smallskip

An intriguing question to be answered is 
``Does the catastrophic vacuum decay in QED really occur?'' 
Although we have discussed the Schwinger effect in QED so far, 
let us change the direction a little here and focus upon a critical behavior in string theory. 
It is well known that there exists an upper critical value of electric field 
in the context of string theory \cite{max1,max2}. This bound obeys from a 
self-consistency condition of string theory. On the other hand, we know 
intimate relations between string theories and gauge theories. 
For example, these are established through the AdS/CFT duality \cite{M,GKP,W}. 
According to this duality, a string theory on an anti-de Sitter (AdS) space 
is equivalent to a conformal field theory (CFT). The most well-studied example is 
the duality between type IIB string theory on ${\rm AdS}_5\times S^5$ and 
the ${\cal N}\! =4$ supersymmetric Yang-Mills theory (SYM) with an $SU(N)$ gauge group 
in four dimensions. 

\smallskip

Thus, one may expect a connection of electric fields between 
the string theory and gauge theory sides through the AdS/CFT duality. 
Then the result in the string-theory side suggests that 
the catastrophic vacuum decay can really occur at least in a class of gauge theories which have gravity duals. 
Thus, to investigate the possibility of the catastrophic vacuum decay in gauge theories much deeper 
beyond the weak-field approximation, 
it would be worth studying the Schwinger effect in the context of the AdS/CFT correspondence \cite{SZ}. 

\smallskip

Another motivation is to argue the Schwinger effect in QCD. 
It may give rise to a new mechanism of a confinement/deconfinement phase transition 
and it would have a connection with the RHIC and LHC experiments, where 
strong electro-magnetic fields and color fields are induced due to the collision of heavy ions. 
It motivates us to study the Schwinger effect in confining gauge theories. 
However, it is not easy to tackle this issue with the standard method in quantum field theories. 
A nice way is to employ a holographic computation by realizing confining gauge theories 
with appropriate D-brane set-ups. In fact, we have considered 
the Schwinger effect in confining gauge theories along this line \cite{SY3,SY4,KSY}. 
The potential analysis \cite{SY3,SY4} has been done for general confining backgrounds \cite{Son1,Son2,Son3} 
by generalizing the procedure in the Coulomb phase\cite{SY2}. 
The production rate has been evaluated numerically in the case of a confining 
D3-brane background \cite{KSY} (which is an example of AdS solitons \cite{HM}).  

\smallskip

The organization of this review is as follows.
Section 2 gives a short review of the world-line instanton method based 
on articles\cite{AAM,Dunne2,Dunne3}. 
This is a method to calculate the production rate and becomes a key ingredient  
in preparing a holographic computation in the subsequent sections. 
In section 3, we consider the Schwinger effect 
in the ${\cal N}\! =4$ SYM. This system is conformal and contains no fundamental matter field. 
Hence, naively, the Schwinger effect cannot be argued. A possible way of overcoming these points 
is to employ the Higgs mechanism. 
We first compute the production rate in the Higgsed ${\cal N}\! =4$ SYM naively 
by using the world-line instanton method. After all, this computation leads to a puzzle 
on the critical value of external electric field. 
Then, to resolve this puzzle, an improved holographic set-up is presented  
by following the seminal work by Semenoff and Zarembo \cite{SZ}. 
The production rate is reconsidered along this line and the puzzle is certainly resolved. 
Section 4 is the main part of this article based on a series of our works \cite{SY3,SY4,KSY} 
and investigates the Schwinger effect in a confining gauge theory 
by employing a confining D3-brane background \cite{HM}. 
A remarkable feature is that there is another kind of critical value of electric field 
due to the existence of the confining phase. 
We have evaluated the production rate numerically and computed new exponents associated with 
the critical electric fields.    
Section 5 is devoted to conclusion and discussion.

\section{The world-line instanton method}

In this section, we will give a short review of the world-line instanton method 
by following the works\cite{AAM,Dunne2,Dunne3}. 

\subsection{The production rate at weak coupling}

There are various methods to calculate the pair production rate of the Schwinger effect in QED.
We shall introduce here one of them, called the world-line instanton method.
An advantage of this method is 
that one can easily add a contribution coming from the Coulomb potential.
In particular, it enables us to compute the production rate at arbitrary coupling.
In the following, we will explain first the computation at weak coupling, 
then the one at arbitrary coupling.

\smallskip

As a simple example, let us first consider a massive scalar QED. 
The action for a massive scalar QED (in the Euclidean signature) is 
\begin{equation}
S=\int \! \dd ^4x\, \left( \frac{1}{4}F_{\mu \nu}^2+|D_{\mu}\phi |^2+m^2 |\phi |^2 \right)
\end{equation}
with the covariant derivative, 
\begin{equation}
D_{\mu}=\partial_{\mu}+ieA_{\mu}+ieA_{\mu}^{\ex}\,.
\end{equation}
Note here that the vector field is divided into 
a dynamical (fluctuating) part $A_\mu$ and an external part $A_{\mu}^{\ex}$\,. 
The path integration is performed only for the dynamical part, not for the external part. 

\smallskip

The production rate $\Gamma$ is given by the imaginary part of the 
vacuum energy density $\varepsilon _0$\,, 
\begin{equation}
\Gamma =2 \, \im \, \varepsilon _0 =-\frac{2}{V_4}\, \im \log \int \! \pd A \, \pd \phi \, {\rm e}^{-S}\,.
\end{equation}
Here $V_4$ is the volume of the four dimensional space. 
By performing the path integral for the scalar field $\phi$, 
the following expression is obtained, 
\begin{equation}
V_4\Gamma =-2 \, \im \log \int \! \pd A \, \exp \left(-\int \! \dd ^4x \, 
\frac{1}{4}F_{\mu \nu}^2-\tr \log \left(-D^2+m^2 \right) \right) \,.
\label{rate1}
\end{equation}
In a small coupling region, the dynamical part of the vector field 
can be ignored. Hence the expression (\ref{rate1}) is approximated as 
\begin{align}
V_4\Gamma &\simeq -2 \, \im \int_0^\infty \! \frac{\dd T}{T}   \int \! \pd x \, \exp 
\left(-\frac{1}{2T}\int_0^{1} \! \dd  \tau \, \dot{x}^2 -\frac{m^2T}{2}
+ie\int_0^1 \! \dd \tau \, A_{\mu}^{\ex}\dot{x}_{\mu} \right)\,. 
\label{10}
\end{align}
Here we have used Schwinger's parametrization 
\[
\log \alpha =- \int _0^\infty  \frac{\dd t}{t}{\rm e}^{-\alpha t}
\]
and the quantum-mechanical path-integral representation. Note that 
$x_\mu (\tau)$ satisfies the periodic boundary condition $x_\mu (0)=x_\mu (1)$\,. 

\smallskip

Next, as for the expression \eqref{10}\,, the $T$-integration is firstly performed, 
and then the integral for $x(\tau)$ is done\footnote{
The integrations in the opposite order lead to the Heisenberg-Eular Lagrangian \cite{HE}.}. 
For later convenience, let us suppose that 
\begin{equation}
m\sqrt{\int_0^1\dd \tau \, \dot{x}^2}\gg 1\,.
\label{weak}
\end{equation}
This condition \eqref{weak} will be identified with the weak-field condition for the 
external electric field, as we will see later. 
It is worth noting that the integration about $T$ can be regarded 
as a modified Bessel function, 
 \[
K_0(x)=\frac{1}{2}\int_0^\infty \frac{\dd t}{t} \exp \left(-t-\frac{x^2}{4t}\right)
\]
and its asymptotic behavior for large $x$ is 
\begin{equation}
K_0(x)\simeq \sqrt{\frac{\pi}{2x}}\, {\rm e}^{-x}\,. 
\end{equation} 
Then, under the condition (\ref{weak}), the expression \eqref{10} is simplified as 
\begin{align}
V_4\Gamma =&-2 \, \im \int \! \pd x \, \frac{1}{m}\sqrt{\frac{2\pi}{T_0}}\exp 
\left(- S_{\text{particle}}\right)\,, \label{rate3}
\end{align}
where $S_{\text{particle}}$ and $T_0$ are defined as 
\begin{align} 
S_{\text{particle}}  &\equiv 
m\,\sqrt{\int_0^1\dd \tau \, \dot{x}^2} -ie\int_0^1 \! \dd \tau \, A_{\mu}^{\ex}\,\dot{x}_{\mu}\,, 
\label{action} \\
T_0 &\equiv \frac{\sqrt{\int \dd \tau \, \dot{x}^2}}{m}\,. 
\end{align}
The above computation is identical with the method of steepest descent for the $T$-integration. 
But, note that the condition \eqref{weak} is not necessary if we do not 
want to use the asymptotic form of the Bessel function at this stage. 

\smallskip

The path integral about $x(\tau)$ is evaluated by the method of steepest descent.
The equation of motion obtained from $S_{\rm particle}$ is 
\begin{equation}
\frac{m\ddot{x}_\mu}{\sqrt{\int \dd \tau \, \dot{x}^2}}=-ieF_{\mu \nu}\dot{x}_\nu\,.
\label{eom}
\end{equation}
Multiplying \eqref{eom} by $\dot{x}_\mu$ and integrating it, 
we find the following relation, 
\begin{equation}
\dot{x}^2\equiv a^2~~~(\mbox{constant})\,.
\end{equation}

\smallskip 

Let us suppose a constant electric field is turned on the $x_1$-direction. 
Then the vector potential is taken as 
\begin{equation}
A_1^{\ex}(x_0)=-iEx_{0} \,.
\end{equation}
The other components are set to be zero. 
Note that the imaginary unit $i$ appears because we are now working in the Euclidean signature. 

\smallskip

With the constant electric field, 
the classical solutions are represented by 
\begin{equation}
x_0=\frac{m}{eE}\cos  \left( \frac{aeE}{m}\tau \right) \,, \qquad
x_1=\frac{m}{eE}\sin  \left( \frac{aeE}{m} \tau \right) \,, \qquad x_2=x_3=0\,.
\label{solution}
\end{equation}
The periodic boundary condition $x_\mu (0)=x_\mu (1 )$ determines $a$ like  
\begin{equation}
\frac{aeE}{m}=2n\pi\,, \qquad n\in \mathbb{N}\,.
\label{a}
\end{equation}
Thus the classical motion has been completely fixed. 
This solution is often called ``instanton,'' just because 
this is a classical solution in the Euclidean signature. 

\smallskip

With the relation \eqref{a}, it is easy to check that the assumption 
\eqref{weak} is nothing but the weak-field condition 
\begin{equation}
E\ll E_{\text{Sch}}\,,
\label{weakfield}
\end{equation}
as noted before.
Furthermore, the assumption is the same as the condition to ensure that 
the steepest-descent method is a good approximation.

\smallskip 

By combining the classical action 
\[
S_\cl ^{(n)}=\frac{\pi m^2 n}{eE}
\]
with one-loop prefactors, 
the production rate is evaluated as 
\begin{equation}
\Gamma =\frac{(eE)^2}{(2\pi)^3}\sum_{n=1}^\infty \frac{(-1)^{n+1}}{n^2}
\exp \left(-\frac{\pi m^2}{eE}n\right)\,.
\label{aamrate}
\end{equation}
The production rate \eqref{aamrate} is the same as the Schwinger formula for the scalar QED.  
This expression is valid at small coupling and under the weak-field condition. 
The method presented here is applicable to a spinor QED as well \cite{Dunne2}.

\smallskip 

We have not explained the one-loop prefactor here. 
For the detail of the derivation, for example, see the works \cite{AAM, Dunne3}.

\subsection{The production rate at arbitrary coupling}

Let us generalize the production rate \eqref{aamrate} at weak coupling to 
the one at arbitrary coupling, though we still suppose the weak-field condition. 

\smallskip 

Before going to the detail, it is worth to see a heuristic argument 
based on the pair production rate of monopole and anti-monopole. 
In the Georgi-Glashow model, it is evaluated as \cite{AM}
\begin{equation}
\Gamma =\frac{(gB)^2}{(2\pi)^3}\exp \left( -\frac{\pi M^2}{gB}+\frac{g^2}{4}\right)\,,  
\label{m-rate}
\end{equation}
where $B$, $M=4\pi m/e^2$ and $g=4\pi/e$ are an external magnetic field, 
a monopole mass and a magnetic charge, respectively.
This result \eqref{m-rate} is valid under the following two conditions:
\[
 g^2 \gg 1\,, \qquad gB \ll M^2\,.
\] 

\smallskip

Let us perform electric-magnetic duality for the above result. 
The production rate is mapped to
\begin{equation}
\Gamma =\frac{(eE)^2}{(2\pi)^3}\exp \left( -\frac{\pi m^2}{eE}+\frac{e^2}{4}\right)\,. \label{conjecture}
\end{equation}
But the validity region is 
\[
e ^2 \gg 1\,, \qquad e E \ll m^2\,,
\]
and the strong coupling region is supposed as opposed to the previous computation. 
This observation indicates the expression \eqref{conjecture} should be valid for arbitrary coupling. 
In fact, this expression can be obtained by a direct computation \cite{AAM}, as we will show below. 

\smallskip 

In the previous subsection, 
the coupling constant has been supposed to be so small that 
the dynamical field $A_\mu$ is ignorable. 
But we will take account of $A_\mu$ at finite coupling here. 
Then the equation \eqref{rate1} is replaced by 
\begin{equation}
V_4\Gamma =-2  \, \im \log \left\langle \exp 
\left[-\tr \, \log \left(-\left(\partial +ieA +ieA^\ex \right)^2+m^2\right)
\right] \right\rangle\,,
\label{286}
\end{equation}
where the expectation value is defined as 
\begin{equation}
\left \langle g(A) \right \rangle \equiv 
\frac{\displaystyle \int \! \pd A \, \exp \left( -\frac{1}{4}\int \! \dd ^4x \, F^2\right) g(A)}
{\displaystyle \int \! \pd A \, \exp \left(-\frac{1}{4}\int \! \dd ^4x \, F^2\right)}\,.
\end{equation}
Furthermore, the equation \eqref{286} can be approximated as
\begin{equation}
V_4\Gamma 
\simeq 2\, \im \, \left \langle \tr \, \log \left(-\left 
(\partial +ieA +ieA^\ex \right)^2+m^2 \right) \right\rangle
\end{equation}
under the weak-field condition. 
This can be regarded as an electron loop expansion in the Feynman diagrams. 

\medskip 

Here we keep the diagrams with a single electron loop and 
drop off the higher-loop contributions.
This selection is a good approximation when the external field is weak enough.
By doing the same calculation as in the previous subsection, 
the production rate is obtained as 
\begin{align}
V_4\Gamma 
&=-2\, \im \int_0^\infty \frac{\dd T}{T} \int \! \pd x \,\exp 
\left(-\frac{1}{2T}\int_0^1\dd \tau \,\dot{x}^2 -\frac{m^2T}{2}+ie\oint A_{\mu}^\ex \, \dd x_\mu \right) \notag \\
& \times \left \langle \exp \left( ie\oint A_{\mu} \, \dd x_\mu \right) \right \rangle \,.
\label{rateall}
\end{align}
In comparison to \eqref{10}, the expression \eqref{rateall} includes the contribution coming from 
the Coulomb interaction. This modification is reflected as 
a $U(1)$ Wilson loop in \eqref{rateall}.

\smallskip

Under the condition \eqref{weak}\,, 
the integration about $T$ leads to
\begin{equation}
V_4\Gamma \simeq -2\, \im \int \! \pd x \, \frac{1}{m}\sqrt{\frac{2\pi}{T_0}} \exp 
\left(-S_{\text{particle}}-\frac{e^2}{8\pi^2}\oint \! \! \oint \frac{\dd x\cdot \dd y}{(x-y)^2}\right)\,,  
\end{equation}
where we have employed the following identity: 
\begin{equation}
\left \langle \exp \left( ie\oint A_{\mu} \, \dd x_\mu \right) \right \rangle
=\exp \left(-\frac{e^2}{8\pi^2}\oint \! \! \oint \frac{\dd x\cdot \dd y}{(x-y)^2} \right)\,. 
\label{id}
\end{equation}
The derivation of \eqref{id} is straightforward. 

\smallskip

Let us next perform the path-integral about $x(\tau)$\,. 
The Coulomb interaction should change classical solutions, as a matter of course.
However, note here that the Coulomb interaction term is invariant under a scale transformation 
and a rotation in the plane on which the instanton is rotating. 
Hence it does not depend on the size of the classical solutions.  
Thus the classical solutions are not changed 
when the Wilson loop is added as a perturbation.

\smallskip 

The classical action is modified as 
\begin{equation}
S_\cl ^{(n)}= \left(\frac{\pi m^2}{eE}-\frac{e^2}{4}\right)n\,.
\end{equation}
Here we have used the formula 
\begin{equation}
-\frac{e^2}{8\pi^2}\oint \! \! \oint \frac{\dd x\cdot \dd y}{(x-y)^2}=n\frac{e^2}{4}\,, 
\label{coulomb}
\end{equation}
where an unphysical divergence has been ignored.

\smallskip 

Recall that the contributions of the Coulomb interaction are small.  
Then the prefactor of the exponential has not been modified.   
As a result, the production rate is given by 
\begin{equation}
\Gamma =\frac{(eE)^2}{(2\pi)^3}\sum_{n=1}^\infty \frac{(-1)^{n+1}}{n^2}
\exp \left(-\left(\frac{\pi m^2}{eE} - \frac{e^2}{4}\right)n\right)\,.
\label{rateall2}
\end{equation}
Thus we have reproduced the conjectured form (\ref{conjecture}) as the $n=1$ case. 

\smallskip

Note that the production rate is not exponentially suppressed any more just after $E$ has reached 
the critical value 
\[
E=E_{\rm c} \equiv \frac{4\pi m^2}{e^3}\,. 
\]
That is, the vacuum becomes unstable catastrophically above this critical value.  
However, the weak-field condition $E\ll E_{\rm Sch}$ is broken as follows: 
\[
 E_{\rm c} = \frac{E_{\rm Sch}}{\alpha_{\rm s}} \qquad 
\Longrightarrow \qquad  E_{\rm c} \sim 137 E_{\rm Sch}\,. 
\]
Hence it is not certain whether the catastrophic vacuum decay really occurs or not. 
It is significant to answer this question. A possible way is to follow a holographic 
computation, as we will introduce in the next section.

\section{The Schwinger effect in the AdS/CFT correspondence}

In this section, we will consider the Schwinger effect in the context of the AdS/CFT correspondence. 
There are many variations of AdS/CFT, but we will concentrate on the most typical example, 
the duality between type IIB superstring on ${\rm AdS}_5\times S^5$ and the 
${\cal N}\! =4$ $SU(N)$ SYM in four dimensions.
However, in order to argue the Schwinger effect, there are three obstacles: 
\begin{enumerate}
\item \quad The ${\cal N}\! =4$ $SU(N)$ SYM is conformal (i.e., all of the fields are massless)\,. 
\item \quad All of the matter fields belong to the adjoint representation.   
\item \quad There is no $U(1)$ gauge field.
\end{enumerate}
As for (3), as a matter of course, one may consider non-abelian Schwinger effects. 
But it is not so easy to study the issue itself. Hence it is better 
to focus upon an abelian Schwinger effect as a first trial. 

\smallskip

Let us first introduce a $U(1)$ gauge field and fundamental matter fields 
by employing the Higgs mechanism. Then one can compute the pair production rate 
of the fundamental matter fields by following the procedure introduced in Sec.\ 2. 
However, we will encounter a puzzle for the critical value of electric field. 
To resolve it, Semenoff and Zarembo proposed a prescription \cite{SZ} 
and presented an improved holographic set-up\footnote{
For an earlier trial, see the work \cite{GSS}. In Ref.\cite{KO}, it has been shown that an electric field
creates a pair production current on a probe brane at zero temperature.
The finite-temperature case is discussed in Refs. \cite{EMS,AFJK}. }. 
Subsection 3.2 will introduce their proposal.  

\subsection{Set-up}

We start from the ${\cal N}$=4 $SU(N+1)$ SYM theory.  
It consists of a gauge field $\hat{A}_{\mu}~(\mu=0,\cdots,3)$\,,  
six real scalar fields $\hat{\Phi}_I~(I=1,\cdots,6)$
and four Weyl fermions $\hat{\Psi}$\,, the hat is attached for later convenience.
All of the fields belong to an adjoint representation of $SU(N+1)$\,.
One should notice here that the ${\cal N}\!=4$ SYM includes neither a $U(1)$ gauge field 
and nor fundamental matters. 

\smallskip

Then, by breaking $SU(N+1)$ to $SU(N)\times U(1)$ with the Higgs mechanism,  
let us introduce a $U(1)$ gauge field. 
The $SU(N+1)$ fields are decomposed into the $SU(N)$ part, 
the $U(1)$ part and non-diagonal parts: 
\begin{equation}
 \hat{A}_{\mu}=
\begin{pmatrix}
A_\mu &\omega _\mu \\
\omega _\mu ^\dagger &a_\mu
\end{pmatrix}\, , \quad 
 \hat{\Phi}_{I}=
\begin{pmatrix}
\Phi_I &\omega _I \\
\omega _I ^\dagger &m\theta _{I}+\phi _I
\end{pmatrix}\, ,\quad
 \hat{\Psi}=
\begin{pmatrix}
\Psi &\chi \\
\chi ^\dagger &\psi
\end{pmatrix}\,. 
\label{decomp}
\end{equation}
Here $A_\mu~[a_{\mu}]$, $\Phi _I~[\phi_I]$, and $\Psi ~[\psi]$ 
are the $SU(N)~[U(1)]$ gauge, the scalar and the fermionic fields, respectively. 
The vacuum expectation value (VEV) of the scalar fields is supposed to be
\[
\langle \hat{\Phi}_{I} \rangle= \diag (0,\cdots ,0, m\theta _{I}) \,, \qquad \sum_{I=1}^6\theta_I^2=1\,.
\]
In terms of D-branes, the Higgs mechanism corresponds to separating a single D3-brane 
from the remaining stuck of $N$ parallel D3-branes. 
The distance between the separated D3-brane and the $N$ D3-branes 
is related to the VEV of the scalar fields.

\smallskip 

As a result, the ${\cal N}$=4 $SU(N+1)$ SYM action
$S^{SU(N+1)}_{\mathcal{N}=4 \,\sym}$ is decomposed like 
\[
S^{SU(N+1)}_{\mathcal{N}=4 \,\sym} = S^{SU(N)}_{\mathcal{N}=4 \,\sym} +
 S^{U(1)}_{\mathcal{N}=4 \,\sym} + S_{\rm W}\,.
\]
Here $S_{\rm W}$ is the action for the W-boson supermultiplet 
\begin{equation}
S_{\text{W}} = \frac{1}{\gym ^2}\int \! \dd ^4x \, \left[ (D_\mu \omega _I)^\dagger D^\mu \omega _I 
+\omega _I^\dagger \left( \Phi _K-m\theta _K \right)^2\omega _I  
 -m^2  \omega _I^\dagger \theta _I \theta _J \omega _J +\cdots \right]\,. \label{W-mul}
\end{equation}
For our purpose, we will concentrate on the complex scalar fields $\omega_I$ (called ``quarks'') 
and drop off the vector field $\omega _\mu$ and the fermionic field $\chi$ 
in the W-boson supermultiplet. 
The higher order interactions are also ignored. 

\smallskip 

A remarkable point is that the covariant derivative $D_{\mu}$ in (\ref{W-mul}) is 
given by 
\[
D_{\mu} = \partial_{\mu} + i a_{\mu} -i A_{\mu}\,.
\]
Thus the scalar fields in the W-boson supermultiplet 
have a finite mass $m$ and couple to the $U(1)$ gauge field $a_{\mu}$ as well as 
the $SU(N)$ gauge field $A_{\mu}$\,. 

\smallskip 

Note here that five scalar fields are massive but 
one scalar field is massless. For example, let us consider the case with 
$\theta_I=(0,0,0,0,0,1)$\,. For $I=1,\ldots,5$\,, mass terms exist in the action \eqref{W-mul}\,, 
but for $I=6$ it vanishes. The massless field is absorbed into the vector field 
$\omega_{\mu}$ as the longitudinal mode so that $\omega_{\mu}$ becomes massive. 

\smallskip

Let us calculate the pair production rate\footnote{
Recall that $\omega_I$\,, $\omega_\mu $ and $\chi$ have the same mass. 
Hence $\omega_I$\,, $\omega_\mu $ and $\chi$ may be produced equally. 
In fact, the exponential factor does not depend on spins, though the prefactors 
are different. 
In addition, as for $\omega_{\mu}$\,, there is another kind of problem called 
the Nielsen-Olsen instability.  When $\omega_\mu $ pairs are created under 
external electric and magnetic fields. See the work \cite{BKR} for the detail 
of this instability in the holographic Schwinger effect.
}  
for $\omega_I$ by using the world-line instanton method. 
In the following, we take the 't~Hooft limit \cite{Hooft} 
where $N \to \infty$ with $\lambda \equiv \gym ^2 N$ fixed. 
The value of $\lambda$ is also supposed to be sufficiently large so as to suppress  
the dynamical part of the $U(1)$ gauge field. Then $a_{\mu}$ can be regarded 
as an external field like 
\begin{equation}
a_1 ^\ex =-iEx_{0}  \qquad \mbox{(the other components are zero)}\,.
\end{equation}
Note that we work in the Euclidean signature to calculate the production rate.

\smallskip 

Then the production rate can be evaluated as\footnote{For the detailed derivation,  
see sec.\ 2 in the work \cite{SY}.
It is basically the same as the derivation of a supersymmetric Wilson loop 
in an earlier work \cite{DGO}, up to prefactors.}  
\begin{align}
V_4\Gamma =& \,-5N \, \im \int \! \pd x \, g[x(\tau)]\exp \left( -m\int _0^1\! \dd \tau \, 
\sqrt{\dot{x}^2}-i\int _0^1 \! \dd \tau \, a_\mu ^\ex \dot{x}_\mu\right) \langle W[x] \rangle\,, 
\label{n=4rate} \\
 & W[x] =  \tr _{SU(N)} \, \exp \left(\int _0^1 \! \dd \tau \, 
\left(  iA_\mu \dot{x}_\mu  +\Phi _K \theta _K \sqrt{\dot{x}^2}\right) \right)\,,
\end{align}
where $g[x(\tau )]$ is a functional of $x(\tau )$ and does not contribute to the exponential factor 
in the production rate. In the derivation of \eqref{n=4rate}, 
we have assumed that quark mass is very heavy, and the external electric field is weak, $E\ll m^2$\,. 

\smallskip

It would be helpful to comment on some differences between \eqref{rate3} and \eqref{n=4rate}.
Since the production rate is proportional to the number of fundamental particles,
$5N$ appears in \eqref{n=4rate} 
(note that one of $\omega_I$ has been eaten by $\omega_{\mu}$). 
The kinetic term for the instanton is changed, but this change does not affect 
the exponential factor in the leading-order computation. 
Note also that  $W[x]$ in \eqref{n=4rate} is a supersymmetric $SU(N)$ Wilson loop \cite{Wilson1,Wilson2}  
because the $SU(N)$ gauge field $A_{\mu}$ has been regarded as a dynamical field. 

\smallskip

Now it is necessary to evaluate the VEV of the Wilson loop. 
One can compute it by examining the area of the minimal surface of a string attaching to the circle. 
The result at strong coupling is given by \cite{BCFM,DGO}  
\[
\langle W[x] \rangle={\rm e}^{\sqrt \lambda}\,,
\]
by supposing that the mass of quarks should be very heavy. 
Then the exponential factor in the production rate is evaluated as     
\begin{equation}
\Gamma \sim {\rm e}^{-S_\cl} =\exp \left( -\frac{\pi m^2}{E}+\sqrt{\lambda} \right)\,. 
\label{n=4rate2}
\end{equation}
This is an expected form because $\sqrt{\lambda}$ is regarded as a coupling constant 
in the large $N$ gauge theory \cite{Wilson1,Wilson2}.
From \eqref{n=4rate2}, one can read off the critical electric field, 
\begin{equation}
E_{\rm c} = \frac{\pi m^2}{\sqrt{\lambda}}\,. \label{39}
\end{equation}
When $E < E_{\rm c}$\,, the pair production is suppressed exponentially. 
When $E > E_{\rm c}$\,, the vacuum becomes unstable catastrophically. 

\smallskip

It may seem that the critical electric field \eqref{39} satisfies the weak-field condition 
because $\lambda$ is supposed to be very large. However, it would be unlikely that 
heavy particles are produced due to the Schwinger effect because 
such a process is severely suppressed. Hence the above computation 
may be unsatisfactory. 

\smallskip

In addition, note that the critical electric field \eqref{39} does not agree with 
the one obtained from the DBI action of a D3-brane sitting near the AdS boundary\footnote{
If the D3-brane is located at the boundary, then the mass $m$ diverges. 
Hence $m$ has become finite by slightly separating the D3-brane from the boundary.},
\begin{equation}
E_{\rm c}^{\rm DBI} = \frac{2\pi m^2}{\sqrt{\lambda}}\,.
\end{equation}
This disagreement  may lead to a puzzle. 

\smallskip 

In the next subsection, we will introduce a nice prescription to resolve 
the above two points simultaneously.

\subsection{Semenoff-Zarembo's prescription}

In the work \cite{SZ}, Semenoff and Zarembo proposed a holographic set-up 
to study the Schwinger effect in the higgsed ${\cal N}\! =4$ SYM.

\smallskip

\begin{figure}[tbp]
\begin{center}
\includegraphics[scale=.4]{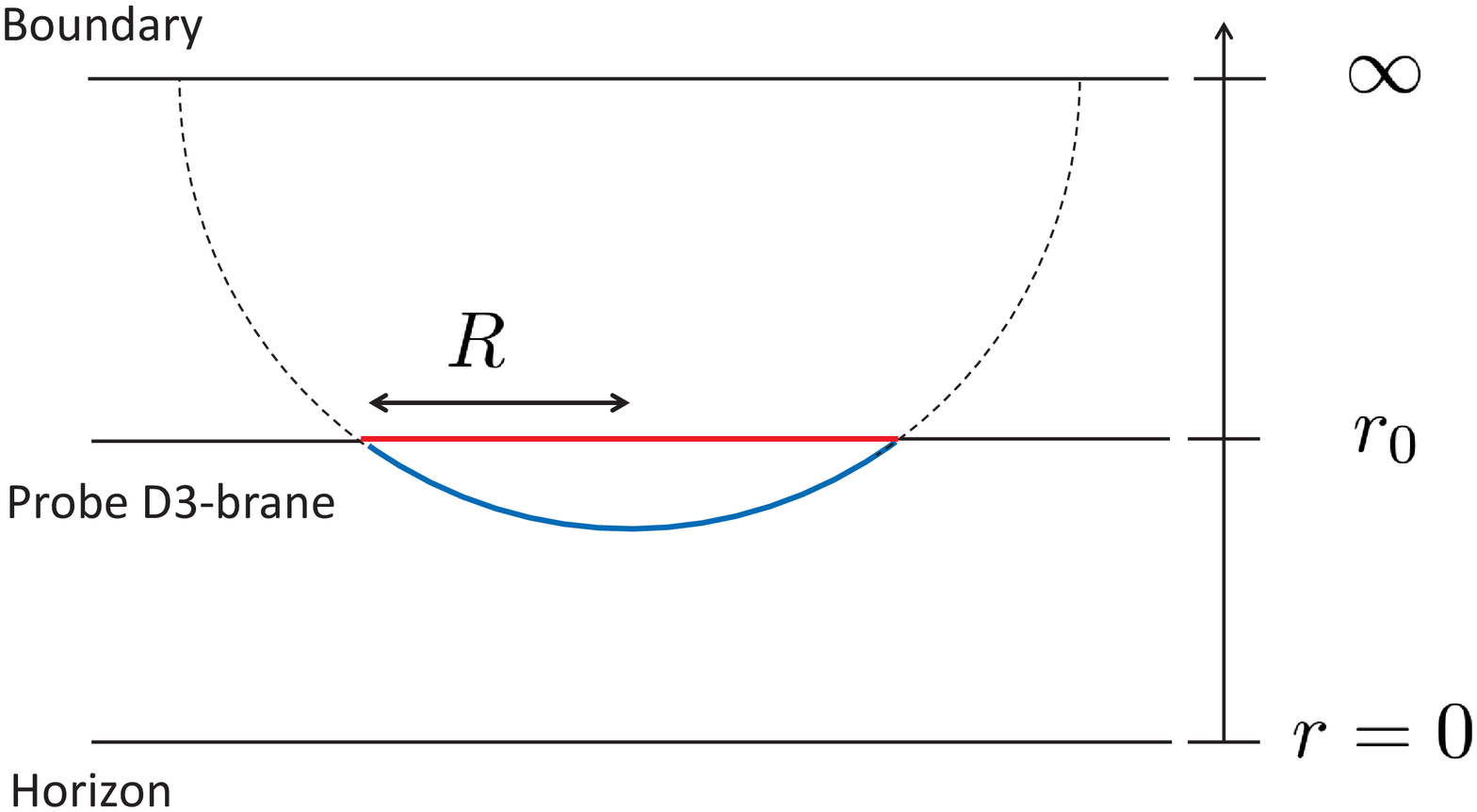}
\end{center}
\vspace*{-0.5cm}
\caption{\footnotesize The location of the probe D3-brane 
and the configuration of the string world-sheet.}
\label{configuration:fig}
\end{figure}

The set-up is the following.
A single D3-brane is separated from the stuck of $N$ parallel D3-branes.  
The near-horizon limit of the stuck of $N$ D3-branes is represented by 
the $\ads_5 \times S^5$ geometry:  
\begin{align}
\dd s^2&=g_{MN}\dd x^M \dd x^N \nonumber \\
&=\frac {r^2}{L^2} \dd x_\mu \dd x^\mu +\frac{L^2}{r^2}\dd r^2 + L^2 \dd \Omega _{5}^2\,,
\end{align}
where $L=\lambda ^{1/4}\sqrt{\alpha '}$ is the curvature radius.  
The coordinates $x^M~(M=0,\ldots,9)$ describe the ten-dimensional spacetime and 
$x^{\mu}~(\mu=0,\ldots,3)$ represent the four-dimensional spacetime 
in which the dual gauge theory lives. The coordinate $r$ is the radial direction of the AdS space.
The horizon of AdS is located at $r=0$ and the conformal boundary is at $r=\infty$\,. 
The separated D3-brane can be treated as a probe brane in the bulk AdS.
A remarkable point is that the probe D3-brane is put at an intermediate position ($r=r_0$) 
between the horizon and the boundary, and it sits at a point on $S^5$\,.
For the configuration, see Fig. \ref{configuration:fig}.

\smallskip

According to the set-up, the mass of a single quark is not infinite any more 
but finite. The quark mass is measured by the energy of a string stretched from the probe D3-brane 
to the horizon like 
\begin{equation}
m=\tf r_0\,.
\end{equation}
Here $\tf=1/2\pi \alpha '$ is the string tension.

\smallskip

Inspiring from the fact that a circular Wilson loop appears in the production rate \eqref{n=4rate}, 
Semenoff and Zarembo proposed a new scenario that 
the production rate may be computed by a circular Wilson loop on the probe D3-brane 
equipped with a constant electric $B$-field. The VEV of the Wilson loop can be evaluated 
by the area of a minimal surface of the fundamental string attaching to the boundary loop.
That is, the exponential factor in the production rate is estimated as 
\begin{equation}
\Gamma  \sim  \exp(-S_{\rm NG} -S_{B_{\it 2}}) \,.
\end{equation}
Here $S_{\rm NG}$ and $S_{B_{\it 2}}$ are the Nambu-Goto (NG) action and the coupling to an NS-NS 2-form.
In the Euclidean signature, these quantities are given by, respectively,  
\begin{align}
S_{\rm NG} &= \tf \int \! \dd ^2 \sigma \, \sqrt{\det G_{\alpha\beta}} \,, \\
S_{B_{\it 2}} &=  -\tf \int \! \dd ^2 \sigma \, B_{MN } \partial _{\tau }x^M \partial _{\sigma }x^N\,.
\end{align}
The string world-sheet is parametrized by $\sigma^{\alpha}=(\tau,\sigma)$\,. 
Then $G_{\alpha \beta}$ is the induced metric and 
$B_{MN}$ is an anti-symmetric two-form flux.
Note again that we work in the Euclidean signature for both the bulk spacetime and the world sheet.

\smallskip 

It would be worth comparing the world-line instanton method and Semenoff-Zarembo's prescription. 
Recall that in the former method the following expression 
\begin{equation}
m\int \! \dd \tau \, \sqrt{\dot{x}^2}-\log \langle W[x]\rangle + i\int \! \dd \tau \, a_\mu ^\ex \dot{x}_\mu 
\label{ins}
\end{equation}
appears in the exponential in \eqref{n=4rate}. In the latter method, the factor \eqref{ins} is replaced by 
$S_{\rm NG} + S_{B_{\2}}$\,. That is, the particle action and the VEV of the Wilson loop are replaced 
by the NG string action $S_{\rm NG}$. Then the NS-NS two-form in $S_{B_2}$ 
is interpreted as an external electric field on the probe 
D3-brane via the following relation,
\[
\tf B_{01}=E\,.
\]

\smallskip

In the 't Hooft limit, the classical analysis is sufficient on the string-theory side.  
To begin with, we construct a classical solution whose boundary is a circular Wilson loop 
on the probe D3-brane by taking the following ansatz (See Fig. \ref{configuration:fig}) :  
\begin{eqnarray}
&& x^0=x(\sigma)\cos \tau \,, \qquad x^1=x(\sigma)\sin \tau \,, 
\qquad x^2 = x^3 =0\,, \nonumber \\
&& r=r(\sigma)\,, \qquad 
x^2+\left( \frac{L^2}{r}\right)^2=R^2+\left( \frac{L^2}{r_0}\right)^2\,, 
\label{szansatz}
\end{eqnarray}
where $R$ is the radius of the circle on the probe D3-brane\footnote{
Note that $x^0=x(\sigma) \cos (n\tau), \, x^1=x(\sigma) \sin (n\tau)$ with $n \geq 2$
satisfy the circular ansatz (\ref{szansatz}). 
However, the contributions of the solutions with $n\geq 2$ are suppressed in comparison to  
the one with $n=1$\,, and hence we will concentrate on the ansatz \eqref{szansatz}. 
}.
The range of $\tau$ is $0\leq \tau <2\pi$\,. 
Note that the coupling to a constant NS-NS 2-form does not change 
the classical equation of motion. Hence the classical solution is given by 
\begin{eqnarray}
x(\sigma) = \frac{1}{\cosh\sigma}\sqrt{R^2 + \left(\frac{L^2}{r_0}\right)^2}\,, 
\qquad r(\sigma) = \frac{L^2}{\tanh\sigma \sqrt{R^2 + ({L^2}/{r_0})^2}}\,. 
\label{cl-sol}
\end{eqnarray}
Here the range of $\sigma$ is taken as $\sigma_0 \leq \sigma <\infty$ and $\sinh \sigma_0 =L^2/Rr_0$\,.

\smallskip 

The next is to argue the boundary condition on the probe D3-brane. 
Note that the NS-NS 2-form is relevant to the boundary condition, 
while it is blind to the equations of motion. 
In the present case, one needs to impose the following mixed 
boundary condition:
\begin{equation}
\sqrt{\det G} \, G^{\alpha  \sigma}g_{MN} \frac{\partial x^M}{\partial \sigma^\alpha}
-B_{MN}\frac{\partial x^M}{\partial \tau} =0\,.
\end{equation}
This boundary condition determines the radius $R$ like 
\begin{equation}
R=\frac{L^2}{r_0}\sqrt{ \left(\frac{E_{\rm c}}{E}\right)^2-1}\,. 
\label{ins:radius}
\end{equation}
Here we have defined $E_{\rm c}$ as
\begin{equation}
E_{\rm c} \equiv \tf \frac{r_0^2}{L^2}=\frac{2\pi m^2}{\sqrt{\lambda}}\,. 
\label{Ec}
\end{equation}
Note that the classical solution does not exist when $E$ is larger than $E_{\rm c}$\,. 
It is also possible to derive the relation \eqref{ins:radius} by taking a variation 
of $S_{\cl}$ with respect to $R$\,, as originally argued in the work \cite{SZ}.

\smallskip 

Finally, by substituting the classical solution \eqref{cl-sol} to the string action,  
the production rate is evaluated as 
\begin{equation}
\Gamma  
\sim  \exp \left[-\frac{\sqrt{\lambda}}{2} 
\left(\! \sqrt{\frac{E_{\rm c}}{E}} - \sqrt{\frac{E}{E_{\rm c}}}\, \right)^2\, \right]
= \exp \left(-\frac{\pi m^2}{E}+\sqrt{\lambda}-\frac{E}{\pi m^2} \right)
\,.
\label{SZrate}
\end{equation}
The first and second terms in the exponential are the same as the expression \eqref{n=4rate2}. 
The third term should be regarded as a correction term and can be ignored 
under the weak-field condition $E \ll m^2$\,. In fact, the classical action vanishes 
when $E=E_{\rm c}$\,. 

\smallskip 

It is worth comparing the result \eqref{Ec} with the critical electric field derived 
from the DBI action of the probe D3-brane. Here we will work in the Lorentzian signature. 
The DBI action in the bulk AdS is given by
\begin{equation}
S _{\rm DBI}=-T_{\text{D3}}\frac{r_0^4}{L^4 }\int \!\dd ^4 x\, 
\sqrt{1-\frac{(2\pi \alpha ')^2L^4}{r_0^4}E^2}\,, 
\label{d3}
\end{equation}
where the D3-brane tension is given by 
\[
T_{\text{D3}}= \frac{1}{\gs (2\pi )^3\alpha ^{\prime 2}}\,.
\]
One can see that the DBI action is ill-defined when $E > {2\pi m^2}/{\sqrt{\lambda}}$\,.
Thus, the critical electric field is given by 
\begin{equation}
E_{\rm c} ^{\rm DBI}= \frac{2\pi m^2}{\sqrt{\lambda}}\,.
\end{equation}
This completely agrees with the result (\ref{Ec})\,.

\smallskip 

As we have seen so far, Semenoff and Zarembo's proposal works well on  
the critical electric field and has resolved the two unsatisfactory points raised 
at the end of sec.\ 3.1.  
Then the production rate \eqref{SZrate} coincides with 
the result \eqref{n=4rate2} computed with the world-line instanton method, 
under the weak-field condition $E \ll m^2$\,. 
Thus it seems likely that there would be no problem for the exponential factor. 

\smallskip 

However, it has not succeeded yet to evaluate the prefactor 
of the exponential with the present holographic set-up. 
The difficulty comes from the fact there is a subtlety in regularizing  
quantum fluctuations around a circular Wilson loop, in comparison to a straight line.  
In fact, some trials have already been done \cite{DGT,SaY,AmMa,KM}, 
but unfortunately none of them has succeeded. 

\smallskip 

Finally, it is useful to comment on some generalizations of Semenoff and Zarembo's work \cite{SZ}. 
One may consider the finite temperature case \cite{BKR,SY2} by following the works \cite{Witten,HS}.  
Bolognesi et.\,al. \cite{BKR} argued the pair creation of monopole and anti-monopole 
in an external magnetic field. Some cases with external 
electro-magnetic fields are also studied \cite{SY}.

\section{The Schwinger effect in a confining gauge theory}

So far, we have studied a holographic description of the Schwinger effect 
in the Coulomb phase. An interesting direction is to consider the Schwinger effect 
in the confining phase by generalizing the previous set-up.  
This section is devoted to the study of the Schwinger effect in a confining gauge theory  
by employing a holographic approach\cite{SY3,SY4,KSY}.

\smallskip

It would be important to motivate the readers to study the Schwinger effect in the confining phase, 
before going to the detail. Recently, QCD in strong external fields has been intensively studied to explain 
experimental data of RHIC and LHC. In these experiments, a extremely strong magnetic field is induced, 
because colliding nuclei have large electric charge and move very fast. 
This strong magnetic field may have a visible effect on the nuclear dynamics. 
Apart from the magnetic field, a strong electric field is also created around the nuclei as well as a strong color field.
Thus, to capture the underlying physics concerned with the experiments,  
it is inevitable to study the QCD dynamics in the presence of strong external fields. 
As a matter of course, the Schwinger effect in the confining phase is included in this direction. 

\smallskip

We will utilize a holographic approach to study the Schwinger effect, 
but one may wonder whether the lattice formulation would be enough to study it. 
In the presence of an external electric field, one encounter a notorious sign problem 
and hence it is quite difficult to use the lattice formulation in general. 
Some researchers may hesitate to borrow a holographic computation. 
But one cannot disguise the fact that there is no definite method to tackle this issue instead of it, at least so far. 
It would be quite useful to capture qualitative pictures of the Schwinger effect in the confining phase 
even by employing the AdS/CFT. In fact, one can see quantitative predictions as well as 
qualitative understanding by pursuing this direction.

\subsection{Set-up}

First of all, let us introduce the set-up to study the Schwinger effect in the confining phase. 
For this purpose, the bulk AdS$_5$ has to be replaced with another confining background 
with a dimensionful parameter. 
There are lots of backgrounds dual to confining gauge theories. 
A simple example is an AdS$_5$ soliton background 
(often called a confining D3-brane background)\footnote{
In this review, we will concentrate on an AdS$_5$ soliton background for simplicity, 
but it would be easy to generalize the present analysis 
to the other confining backgrounds by following the work\cite{SY4}.}.

\smallskip

The metric of the AdS$_5$ soliton background is given by 
\begin{align}
\dd s^2&=\frac{L^2} {z^2}\left[ -(\dd x^0)^2+\sum _{i=1}^{2}(\dd x^i )^2+
f(z)(\dd x^3 )^2+\frac{\dd z^2}{f(z)}\right] + L^2 \dd \Omega _{5}^2\,, \notag \\
&   f(z)=1-\left( \frac{z}{z_{\rm t}}\right)^{4}\,. \label{metric} 
\end{align}
Now the radial direction is described by $z$ and a scalar function $f(z)$ contains 
a constant parameter $z_{\rm t}$\,. Then the geometry is cut off at $z=z_{\rm t}$ and 
the conformal boundary is at $z=0$\,. 
The parameter $z_{\rm t}$ has the dimension of length and the inverse has the dimension 
of energy. Hence $1/z_{\rm t}$ gives rise to the confining string tension on the dual gauge-theory side. 
As $z_{\rm t} \to \infty$\,, $f(z) \to 1$ and the usual AdS$_5$ geometry is reproduced. 

\smallskip 

As another point, the $x^3$-direction is compactified on a circle $S^1$ with the radius $R = \pi z_{\rm t}$\,. 
Hence the limit $z_{\rm t} \to \infty$ corresponds to the decompatification limit $R\to \infty$\,. 
Anyway, the dual gauge theory lives on $\mathbb{R}^{1,2}\times S^1$ rather than $\mathbb{R}^{1,3}$\,. 
Thus the dual gauge theory cannot be regarded as the usual confining gauge theory in $1+3$ dimensions. 
Still, however, it would be possible to extract qualitative behaviors by examining the AdS$_5$ soliton background. 
The results obtained here would be a key ingredient towards the study in the real QCD.

\subsection{Potential analysis}

Let us first study the Schwinger effect in the confining phase from the 
viewpoint of potential analysis. 
Naively, the Schwinger effect could  be argued with the modified potential
\begin{equation}
V_{\rm tot}(x)=2m +V(x)-Ex + \sigma_{\rm st} x\,. 
\end{equation}
Here the modification is that a confining potential with the string tension $\sigma_{ \rm st}$ is added 
as well as the usual potential analysis. An important observation is that 
the string tension $\sigma _{\rm st}$ and the electric field $E$ compete with each other. 
When $\sigma_{\rm st} > E$\,, the potential diverges as $x \to \infty$ and hence 
the Schwinger effect would not occur. On the other hand, when $\sigma_{\rm st} < E$\,, 
the situation is not so different from the Coulomb phase and 
the qualitative behavior would be almost the same as the previous.   

\medskip 

In the holographic set-up, it is necessary to evaluate the VEV of a rectangular 
Wilson loop for the quark anti-quark potential. 
It can be computed with the minimal surface of a string attaching to 
the loop on the probe D3-brane embedded in the ${\rm AdS}_5$ soliton background. 


\smallskip

\begin{figure}[tbp]
\begin{center}
\includegraphics[scale=.6]{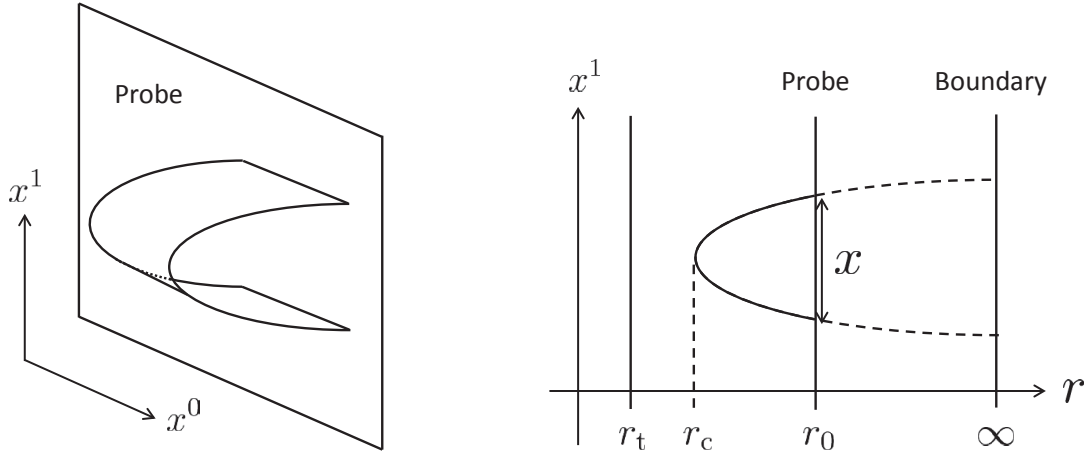}
\vspace*{-0.5cm}
\caption{\footnotesize The configuration of the string world-sheet for the quark and anti-quark potential. }
\label{AdS-soliton:fig}
\end{center}
\end{figure}

In this subsection, we will work with 
\[
r=\frac{L^2}{z}
\]
as the radial coordinate, following the notation of our article\cite{SY3}. 
Hereafter, we move to the Euclidean signature.
The probe D3-brane is supposed to be at $r=r_0$\,.
We impose the following ansatz for the rectangular Wilson loop (See Fig. \ref{AdS-soliton:fig}):
\begin{equation}
x^0=\tau \,, \quad x^1=\sigma \,, \quad r=r(\sigma)\,.
\end{equation}
The NG part of the string action is 
\begin{equation}
{\cal L}=\sqrt{\det G} =\sqrt{\frac{1}{1-\rt^4/r^4}\left(\frac{\dd r}{\dd \sigma}\right)^2+\frac{r^4}{L^4}}\,. 
\label{L}
\end{equation}
Since the Lagrangian does not depend on $\sigma$ explicitly, the Hamiltonian 
\begin{equation}
\frac{\partial \mathcal{L}}{\partial (\partial_{\sigma}r)}\partial_{\sigma}r  - \mathcal{L}\,
\end{equation}
is conserved. When we impose the boundary condition 
at the tip of the minimal surface like 
\begin{equation}
\frac{\dd r}{\dd \sigma} =0\,, \qquad  r=\rc \quad (\rt < \rc < r_0)\,,  
\label{bc}
\end{equation}
the conserved quantity becomes
\begin{equation}
\frac{r^4}{\displaystyle \sqrt{\frac{1}{1-\rt^4/r^4}\left(\frac{\dd r}{\dd \sigma}\right)^2+\frac{r^4}{L^4}} }
= \mbox{const.} \equiv
\rc ^2L^2\,. \label{conserved}
\end{equation}

\smallskip 

By rewriting the conserved quantity, we obtain 
\begin{equation}
\frac{\dd r}{\dd \sigma}=\frac{1}{L^2}\sqrt{(r^4-\rt^4)\left(\frac{r^4}{\rc ^4}-1\right)}\,. \label{diff}
\end{equation}
By integrating the expression \eqref{diff}, the distance between a quark and an anti-quark, $x$, is given by 
\begin{equation}
x=\frac{2L^2}{r_0a}\int_1^{1/a} \! \! \frac{\dd y}{\sqrt{(y^4-1)(y^4-(b /a)^4)}}\,, 
\end{equation}
where we have defined dimensionless quantities as 
\[
y\equiv \frac{r}{\rc}\,, \qquad  a\equiv \frac{\rc}{r_0}\,, \qquad b\equiv \frac{\rt}{r_0}\,. 
\]
The sum of the potential energy and static energy is evaluated as
\begin{align}
V_{\rm PE+SE}=2\tf\int^{x/2}_0\! \dd \sigma \,\mathcal{L} 
=2\tf r_0 a\int_{1}^{1/a} \!\dd y \, \frac{y^4}{\sqrt{(y^4-1)(y^4-(b/a)^4)}}\,. 
\end{align}
Here $x$ is a function of $a$\,, and hence the potential is a function of $x$ thorough $a$\,.

\smallskip

For large $x$ limit (i.e. $a\to b$ limit), the sum of the potential energy and static energy behaves as
\begin{align}
V_{\rm PE+SE}
=\tf  \left( \frac{r_0}{L}\right) ^2b^2x+2\tf r_0b \left( \frac{1}{b}-1\right)\,. 
\end{align}
The first term is the quark and anti-quark potential with a confining string tension 
\[
\sigma_{\rm st} =\tf \left( \frac{\rt}{L}\right) ^2\,,
\]
and the second term is static mass of quark and anti-quark
\[
2 \tf (r_0-\rt ) = 2m_{\rm W}\,.
\] 

\smallskip

By including the energy coming from the external electric field, the total potential is rewritten 
into an integral representation,  
\begin{align}
V_{\rm tot} &=V_{\rm PE+SE}-Ex \notag \\
&=2\tf r_0a \int_{1}^{1/a} \!\dd y \, \frac{y^4}{\sqrt{(y^4-1)(y^4-(b/a)^4)}} 
\notag \\
&\qquad 
-\frac{2\tf r_0\alpha }{a}\int_1^{1/a} \! \! \frac{\dd y}{\sqrt{(y^4-1)(y^4-(b /a)^4)}}\,, 
\end{align}
where we have introduced a dimensionless electric field $\alpha$
\begin{eqnarray}
\alpha \equiv \frac{E}{E_{\rm c}}\,, \qquad E_{\rm c} \equiv \tf \frac{r_0^2}{L^2}\,. 
\label{alpha}
\end{eqnarray}
$E_{\rm c}$ is the critical electric field obtained from the DBI action.
In the following, we define a dimensionless electric field $\alpha$ normalized by $E_{\rm c}$\,.

\smallskip 

\begin{figure}[tbp]
\begin{center}
\includegraphics[scale=.35]{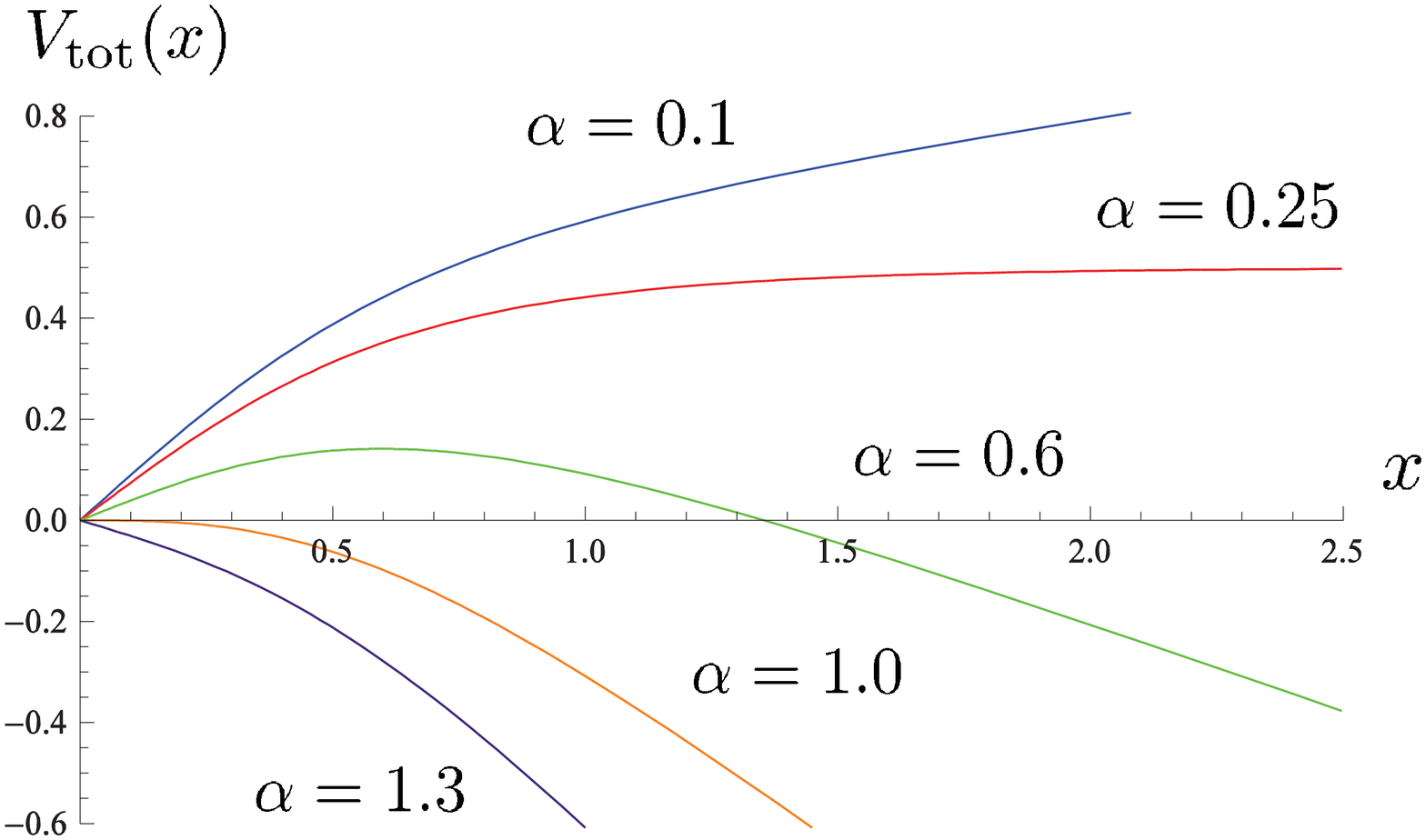}
\vspace*{-0.8cm}
\caption{\footnotesize The plots of the total potential with $b=0.5$ and 
$2L^2/r_0=2\tf r_0=1$\,. 
The blue line is for $\alpha=0.1$\,. There is no zero other than the origin 
and hence the Schwinger effect does not occur. The red line is for $\alpha=0.25$\,. 
This value corresponds to $b^2 =0.25$ and 
the potential becomes flat as $x\to \infty$\,. 
For the values of $\alpha$ between $0.25$ and $1.0$\,, the potential barrier is formed  
and the Schwinger effect can occur as a tunneling process. 
When $\alpha=1.0$ (the orange line), the barrier just vanishes 
and the system becomes unstable catastrophically. 
}
\label{D3-plots:fig}
\end{center}
\end{figure}

The total potential is plotted in Fig.\,\ref{D3-plots:fig} as a function of $x$\,.
Figure \ref{D3-plots:fig} shows the existence of two critical electric fields.
The first one is 
\begin{equation}
E=E_{\rm s} \equiv \sigma _{\rm st} \qquad (\alpha =b^2)\,,
\end{equation}
and the second is 
\begin{equation}
E=E_{\rm c} \qquad (\alpha =1)\,.
\end{equation}
When the electric field is smaller than the confining string tension, 
the pair production is prohibited because the potential does not dump at infinity. 
When the electric field is larger than it, 
it is allowed as a tunneling process.
Thus, at the first critical electric field, a confinement/deconfinement transition starts. 
As a side remark, a similar behavior of the phenomenological potential 
in a confining theory has been argued by using a lattice formulation \cite{lattice}. 

\smallskip

The second critical electric field is the same as the critical electric field obtained 
in the Coulomb phase. That is, the vacuum becomes unstable catastrophically 
above this value of the electric field. Note that this qualitative behavior 
can also be understood analytically. For the detail, see the works \cite{SY2,SY3,SY4}.

%

\smallskip 

Here we are mainly concerned with the long-distance behavior of the potential. 
But it would be worth commenting on the short-distance behavior. 
As shown in Fig.\,\ref{D3-plots:fig}, the potential becomes zero at the origin, 
in comparison to the usual potential analysis. This is just because 
the Coulomb potential is modified on the probe D3-brane 
sitting at an intermediate position in the bulk. 
The modified Coulomb potential was originally argued by Kabat and Lifschytz \cite{KL}.

\subsection{The production rate}

In this subsection, we will compute the production rate numerically  
and examine two critical electric fields. 
Both of the resulting critical values agree with the ones obtained in the potential analysis \cite{SY3}. 
Furthermore, we will introduce new exponents associated with the critical behaviors 
and evaluate the numerical values. 

\smallskip

To compute the production rate, 
one has to evaluate the VEV of a circular Wilson loop on the probe D3-brane.  
Then the D3-brane is placed at an intermediate position between $z=z_{\rm t}$ and $z=0$\,. 

\smallskip

\begin{figure}[tbp]
\begin{center}
\includegraphics[scale=.6]{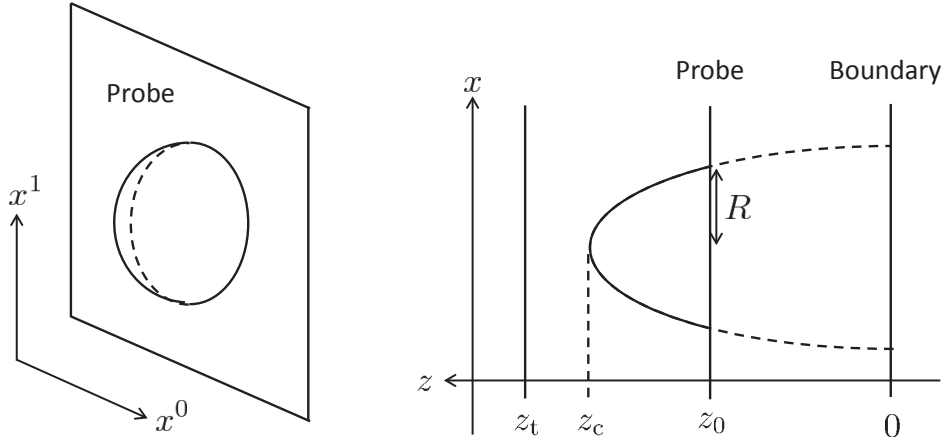}
\end{center}
\vspace*{-0.5cm}
\caption{The configuration of the  string world-sheet for the Schwinger pair production.} 
\label{config-circle:fig}
\end{figure}

The first is to construct a classical string solution attaching to 
the circular Wilson loop. Let us suppose the following ansatz: 
\begin{equation}
x^0=x(\sigma)\cos \tau \,, \qquad x^1=x(\sigma)\sin \tau \,, \qquad z=z(\sigma)\,. 
\label{target}
\end{equation}
The other components are set to be zero. 
Suppose that the world-sheet coordinates $(\tau,\sigma)$ are restricted 
to the following range: 
\[
0 \leq \tau < 2\pi\,, \qquad 0 \leq \sigma \leq \sigma_0\,. 
\] 
Then boundary conditions for $x(\sigma)$ and $z(\sigma)$ are imposed like 
\begin{align}
x(0) =0\,, \qquad x(\sigma_0) = R\,,  \qquad 
z(0) =z_{\rm c}\,, \qquad  z(\sigma_0) = z_0\,. \label{bdc}
\end{align}
The configuration is depicted in Fig.\,\ref{config-circle:fig}. 

\smallskip

In addition, a constant field $B_\2$ is set to be 
\[
\tf B_{01} \equiv E\,.
\]
This $B_\2$ field induces an external electric field on the probe D3-brane.  
Then the resulting NG action and the coupling to $B_\2$ are given by, respectively, 
\begin{align}
S_{\rm NG} &= 2\pi L^2 \tf \int_0^R \! \dd x \, \frac{x}{z^2}\sqrt{1+\frac{z'^2}{f(z)}}\,, \\ 
S_{B_{\it 2}} &= -2\pi \tf B_{01} \int_0^R \! \dd x\, x =-\pi E R^2\,.
\label{actions}
\end{align}
Here we take that $\sigma = x(\sigma)$ with a diffeomorphism invariance.
The prime denotes the derivative with respect to $x$\,. 

\smallskip 
.
Then the equation of motion for $z(x)$ is obtained as 
\begin{equation}
  z'+\frac{2x f(z)}{z} + x z'' - \frac{x z'^2}{2f(z)} \frac{\dd f}{\dd z}(z) + \frac{z'^3}{f(z)}+\frac{2x z'^2}{z}=0\,.
  \label{eom.z}
\end{equation}
In the presence of the $B_\2$ field, the boundary condition on the probe D3-brane 
becomes the mixed boundary condition: 
\begin{equation}
z'=-\sqrt{f(z)\left( \frac{1}{\alpha^2}-1\right)} \,\Biggr|_{z=z_0}\,. 
\label{mixed}
\end{equation}
Here $\alpha$ is a dimensionless electric field defined in (\ref{alpha})\,.

\smallskip

It is difficult to solve analytically the differential equation (\ref{eom.z}) 
with the boundary condition (\ref{mixed})\,. 
Hence let us solve it numerically and study the behavior of $z(x)$\,.
As a result, we will show the existence of two critical electric fields 
and the consistency of the values with the ones obtained from the potential analysis in subsec.\ 4.2.
Hereafter, we take that 
\[
\lambda =100 \quad (2\pi \tf L^2=10) \,,
\]
to validate the holographic description (i.e., large $\lambda$).

\smallskip

\begin{figure}[htbp]
\[
\begin{array}{cc}
\includegraphics[scale=.35]{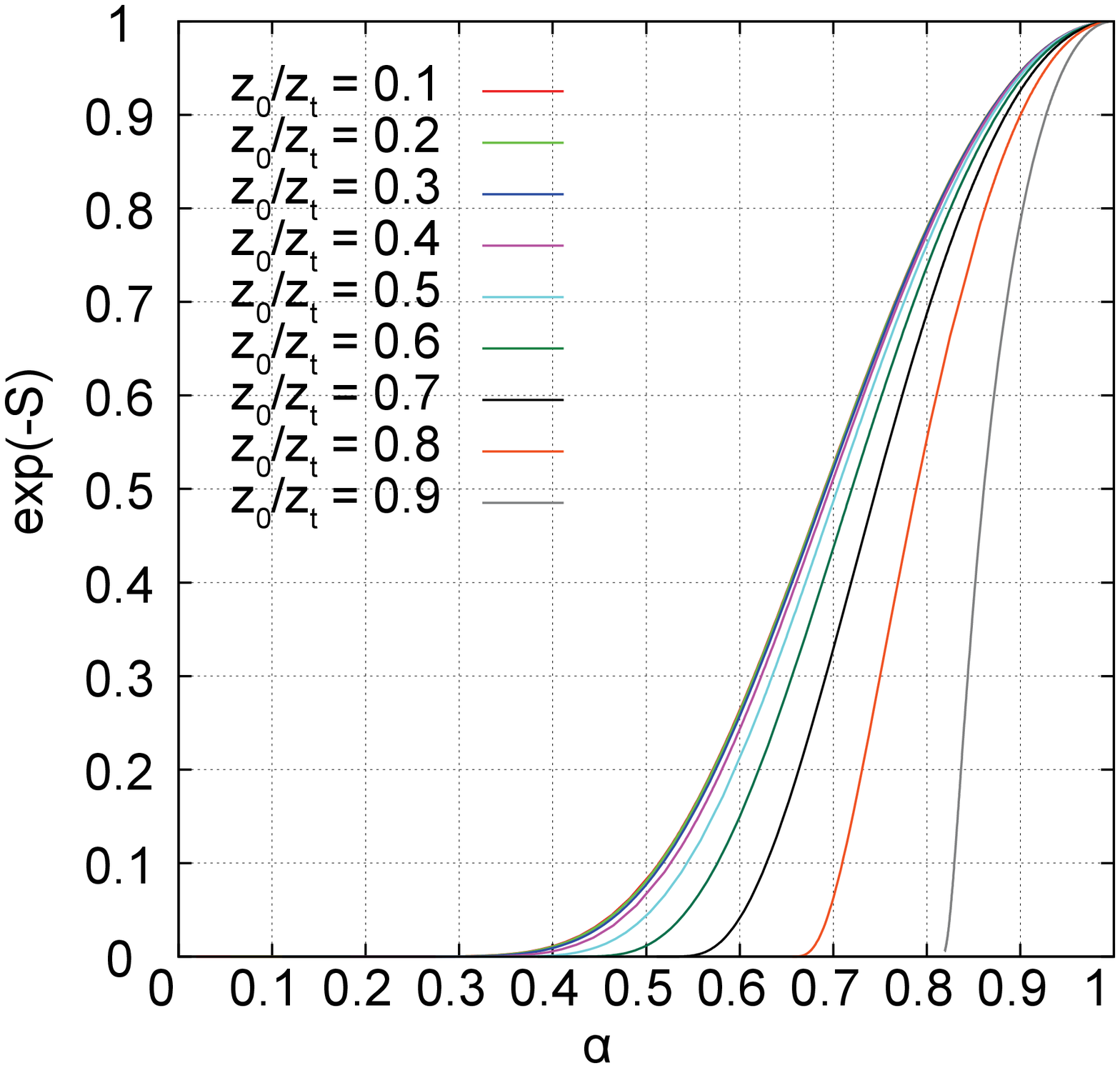} &
\includegraphics[scale=.35]{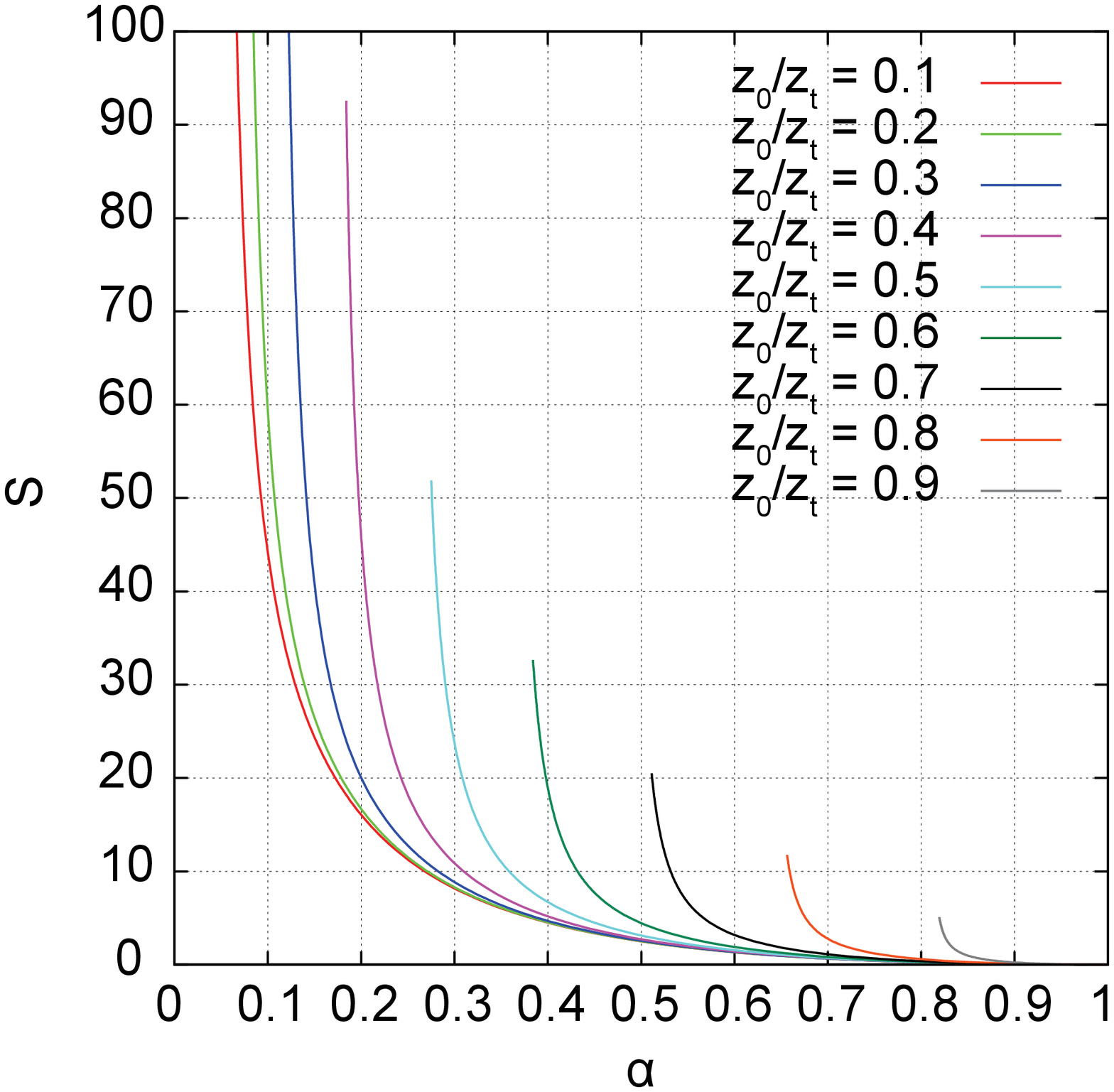} \\ 
\vspace*{0.3cm} \mbox{\footnotesize (a) \quad the exp factor} & 
\mbox{\footnotesize (b) \quad the classical action}
\end{array}
\]
\vspace{-1cm}
\caption{The plots of the exponential factor and the classical action.
} 
\label{rate:fig}
\end{figure}

For some values of $z_0/z_{\rm t}$\,, numerical results are shown in Fig.\,\ref{rate:fig}. 
The left figure shows exponential suppression factors, and 
the right one is plots of the classical action. 
Typically, the suppression factors tend to vanish below certain values of $\alpha$ 
(i.e., $E=E_{\rm c}$)\,, depending on values of $z_0/z_{\rm t}$\,. 
This is the same $E$-dependence as in the Coulomb phase. 
In the case of the confining phase, one can see a new behavior that the classical action diverges  
(i.e. the exponential factor vanishes) at the value $E=E_{\rm s}$\,. 
This result indicates that the Schwinger effect does not occur when $E < E_{\rm s}$\,. 
Note that the value of $E_{\rm s}$ agrees with the potential analysis as well. 
As a matter of course, the value of $E_{\rm s}$ depends on $z_0/z_{\rm t}$\,, 
and $E_{\rm s}$ becomes zero as $z_{\rm t} \to \infty$\,. 

\smallskip 

In total, the two critical electric fields $E_{\rm s}$ and $E_{\rm c}$ have been seen  
from Fig.\,\ref{rate:fig}. These are the same as the ones obtained by the potential analysis in subsec.\ 4.2, 
\begin{equation}
  E_{\rm c} \equiv \tf \frac{L^2}{z_0^2}\,, \qquad  
  E_{\rm s} \equiv \tf \frac{L^2}{z_{\rm t}^2}\,.
\end{equation}
This agreement supports that our numerical results of the production rate 
are consistent with the previous potential analysis. 

\smallskip 

\subsection*{The critical behaviors}

Let us examine the critical behaviors of the classical action numerically. 
We first argue the behavior of the classical action around $E=E_{\rm s}$\,, 
by focusing upon the singular behavior. 
Then the critical behavior around $E=E_{\rm c}$ is discussed numerically and analytically. 
For both critical behaviors, we will introduce new exponents and determine the numerical values.

\subsubsection*{i) the critical behavior around $E=E_{\rm s}$}

\begin{figure}[tbp]
\vspace*{-0.5cm}
\[
\begin{array}{cc}
\includegraphics[scale=.35]{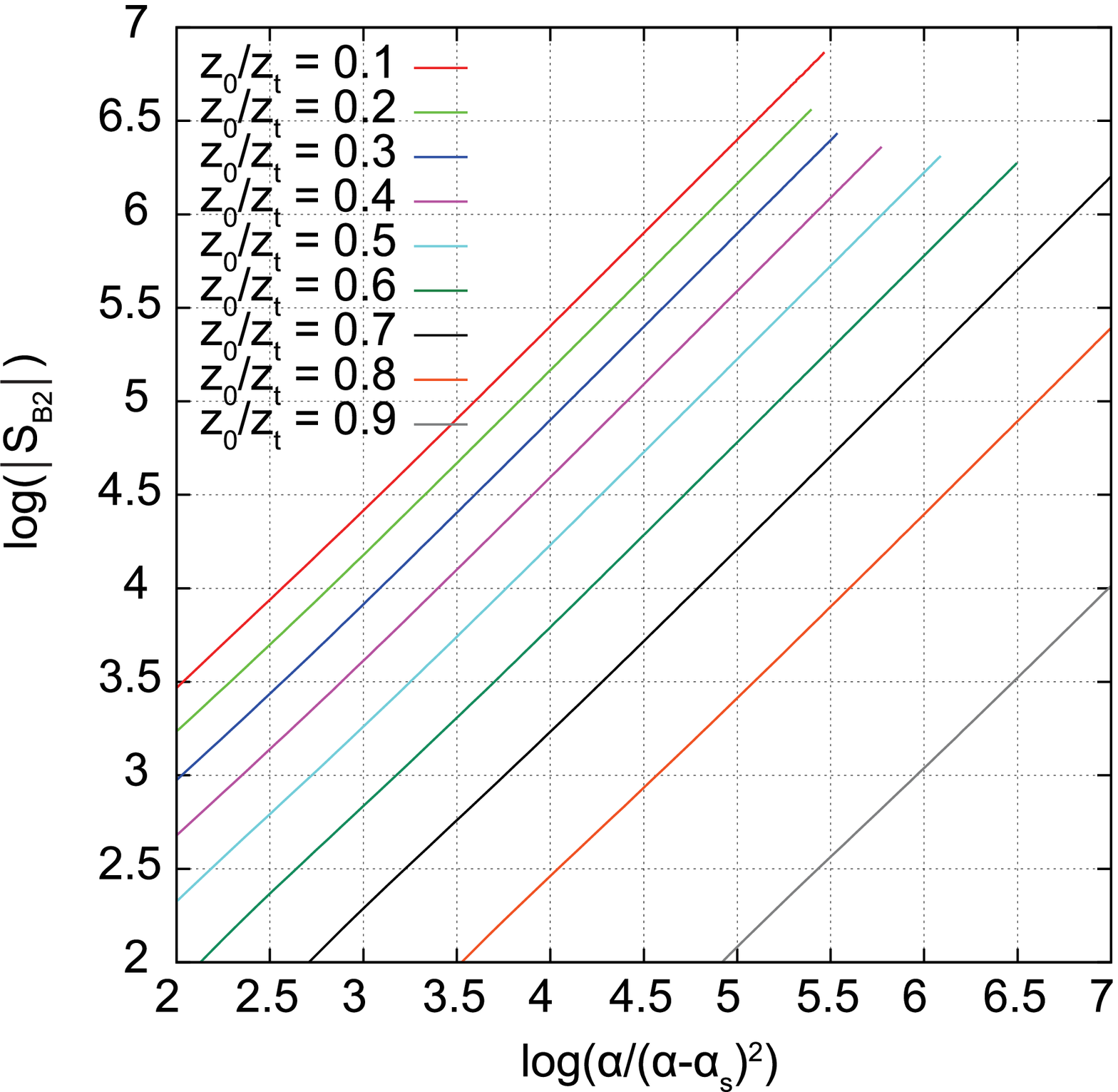} & 
\includegraphics[scale=.33]{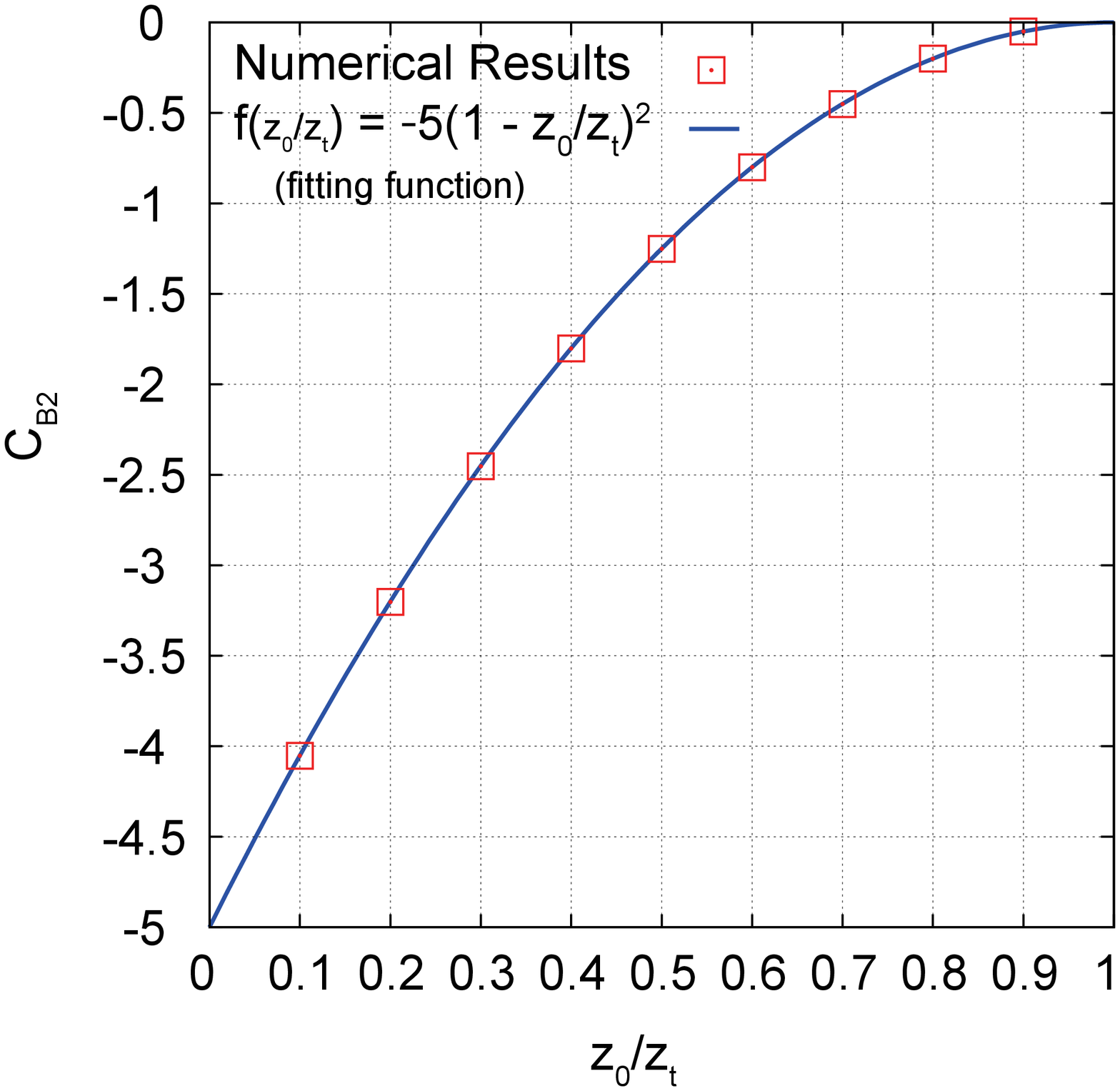} 
\vspace*{0.1cm} 
\\ 
\mbox{\footnotesize (a) \quad $S_{B_\2}$} & 
\mbox{\footnotesize (b) \quad $C_{B_\2}(\alpha_{\rm s})$} 
\end{array}
\]
\vspace*{-0.5cm}
\caption{\footnotesize The behaviors of $S_{B_\2}$ and $C_{B_\2}(\alpha_{\rm s})$ near $E=E_{\rm s}$\,. 
\label{B2:fig}}
\end{figure}

Let us consider the behavior near $E=E_{\rm s}$\,.
The log-log plot of $S_{B_\2}$ shown in Fig.\,\ref{B2:fig} (a) indicates that 
the $\alpha$-dependence of $S_{B_\2}$ may be described by
\begin{equation}
S_{B_\2} = \frac{C_{B_\2}(\alpha_{\rm s})\, \alpha}{(\alpha-\alpha_{\rm s})^2} + \mbox{the regular}\,, 
\qquad \alpha _{\rm s} \equiv \frac{E_{\rm s}}{E_{\rm c}}\,. 
\label{asym} 
\end{equation}
Figure \ref{B2:fig} (b) shows that the coefficient $C_{B_\2}(\alpha_{\rm s})$ is well approximated by
\begin{equation}
C_{B_\2}(\alpha_{\rm s}) = -\frac{\sqrt{\lambda}}{2}(1- \sqrt{\alpha_{\rm s}}\,)^2\,. 
\end{equation}
As a result, by combining (\ref{actions}) and (\ref{asym}), 
the behavior of the Wilson loop radius $R$ near $E=E_{\rm s}$ is evaluated as  
\begin{equation}
R = \frac{(1-\sqrt{\alpha _{\rm s}}\,)z_0}{\alpha -\alpha _{\rm s}}+ \mbox{the regular}\,. 
\end{equation}
This result means that $R$ tends to diverge as $E \to E_{\rm s}+0$\,. 

\smallskip

\begin{figure}[tbp]
\[
\begin{array}{cc}
\includegraphics[scale=.3]{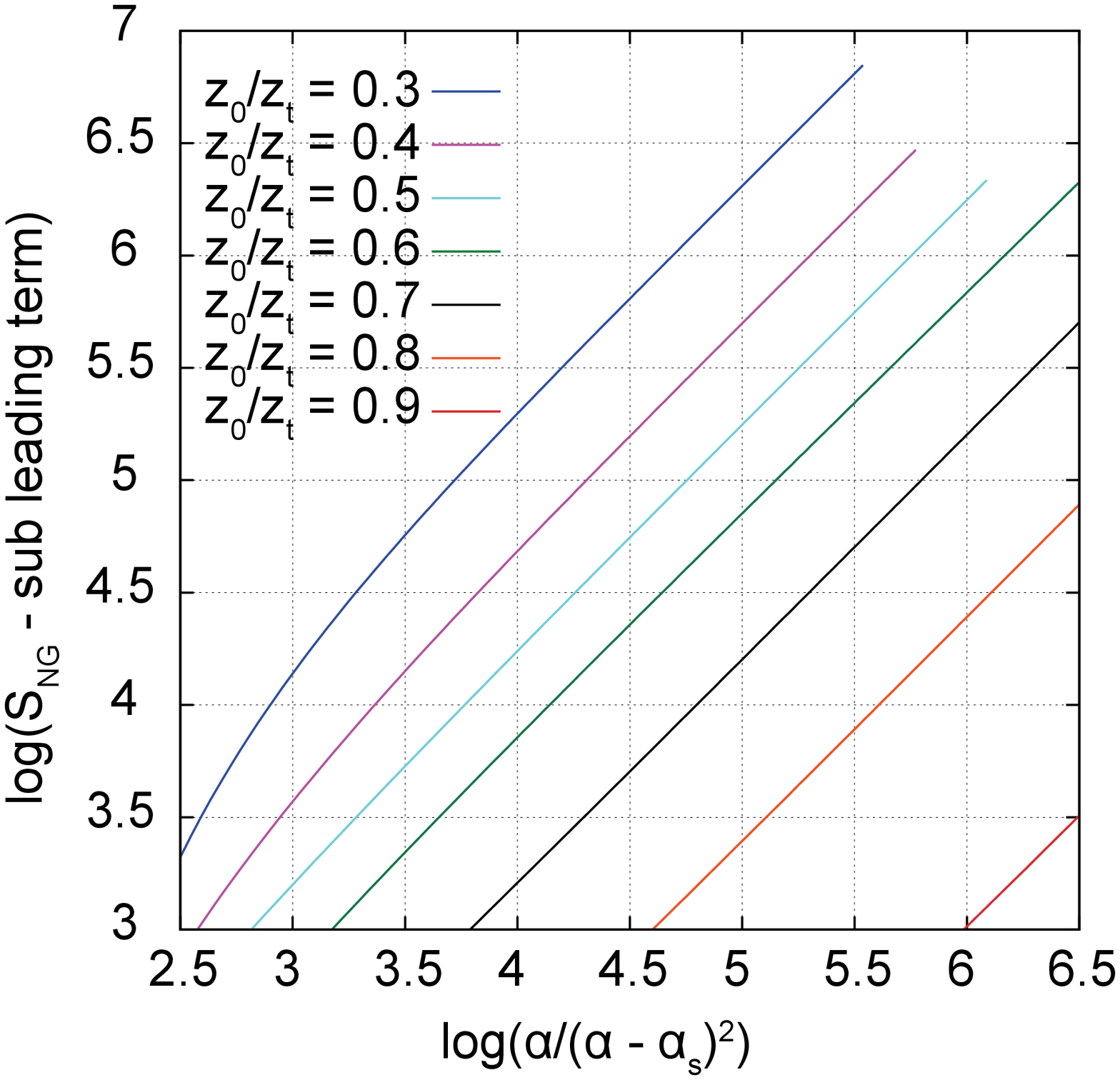} & 
\includegraphics[scale=.3]{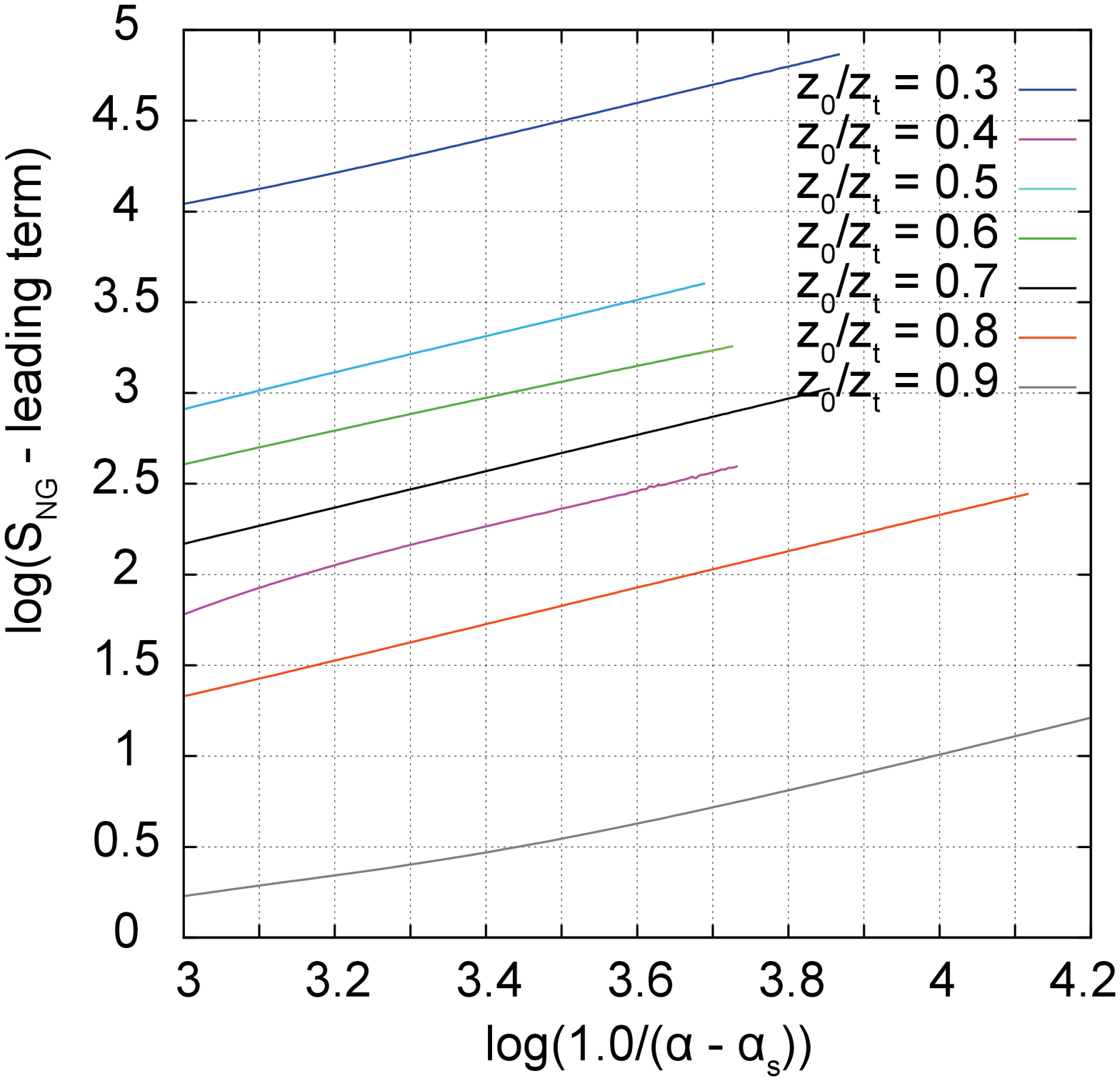} 
\vspace*{-0.2cm} 
\\ 
\mbox{\footnotesize (a)~~the leading term} & 
\mbox{\footnotesize (b)~~the sub-leading term} 
\end{array}
\]
\vspace*{-0.3cm}
\caption{\footnotesize The leading and sub-leading terms of $S_{\rm NG}$ near $E=E_{\rm s}$\,.}
\label{NG:fig}
\end{figure}

The next is to examine the NG action. The log-log plot shown in Fig.\,\ref{NG:fig} (a) 
indicates that the $\alpha$-dependence of the NG action is approximated by
\begin{equation}
S_{\rm NG} = \frac{C_{\rm NG}(\alpha_{\rm s})\, \alpha}{(\alpha-\alpha_{\rm s})^2} 
+ \frac{D_{\rm NG}(\alpha_{\rm s})}{\alpha-\alpha_{\rm s}} 
+ \mbox{the regular}\,. 
\end{equation}
Here $C_{\rm NG}(\alpha_{\rm s})$ and $D_{\rm NG}(\alpha_{\rm s})$ are 
scalar functions to be determined. 

\smallskip

We should comment on the validity of our numerical results. 
The plots in Fig.\,\ref{NG:fig} would be valid for $\alpha_{\rm s} > 0.1$ 
because of the limitation of numerical precision.
If the accuracy of numerical analysis could be increased, then the lines in Fig.\,\ref{NG:fig} 
are extended more to the right-hand side.

\smallskip

In total, the total classical action $S = S_{\rm NG} + S_{B_\2}$ is given by
\begin{equation}
S = \frac{C(\alpha_{\rm s})\,\alpha }{(\alpha -\alpha _{\rm s})^2}+\frac{D(\alpha_{\rm s})}{\alpha -\alpha _{\rm s}} 
+ \mbox{the regular}\,,  
\end{equation}
where we have defined new quantities as 
\[
C(\alpha_{\rm s}) \equiv C_{\rm NG}(\alpha_{\rm s}) + C_{B_\2}(\alpha_{\rm s})\,, \quad 
D(\alpha_{\rm s}) \equiv D_{\rm NG}(\alpha_{\rm s})\,. 
\]
Note that $C(\alpha_{\rm s})$ is nonzero for $\alpha _{\rm s} \neq 0$\,.
Therefore, in the deconfining limit $\alpha_{\rm s} \to 0$\,, the total action $S$ becomes
\begin{equation}
S = \frac{C(0)+D(0)}{\alpha } + \mbox{the regular}\,.    
\end{equation}
Then the following relation should be satisfied, 
\[
C(0)+D(0)= \frac{\sqrt{\lambda}}{2}\,,
\] 
However, we have not confirmed this point yet because of the limitation of numerical analysis 
as mentioned in the last paragraph. 

\smallskip

It would be nice to consider the meaning of the power of the singularity. 
It is associated with the critical behavior and hence may be interpreted as 
a new critical exponent. Suppose the following divergent form around $E=E_{\rm s}$~: 
\begin{equation}
S = A(\alpha_{\rm s}) (\alpha -\alpha _{\rm s})^{-\gamma_{\rm s}}+\cdots \,.
\end{equation}
Now $\gamma_{\rm s}$ is a new exponent. Our numerical computation indicates that 
\[
\gamma_{\rm s}=2\,.
\]
This may be regarded as a non-trivial prediction to the real QCD 
if this value could be interpolated to QCD via the universality argument. 
In this sense, it would be significant to check the universality of this exponent 
for general confining backgrounds.

\subsubsection*{ii) the critical behavior around $E=E_{\rm c}$}

\begin{figure}[tbp]
\[
\begin{array}{cc}
\includegraphics[scale=.3]{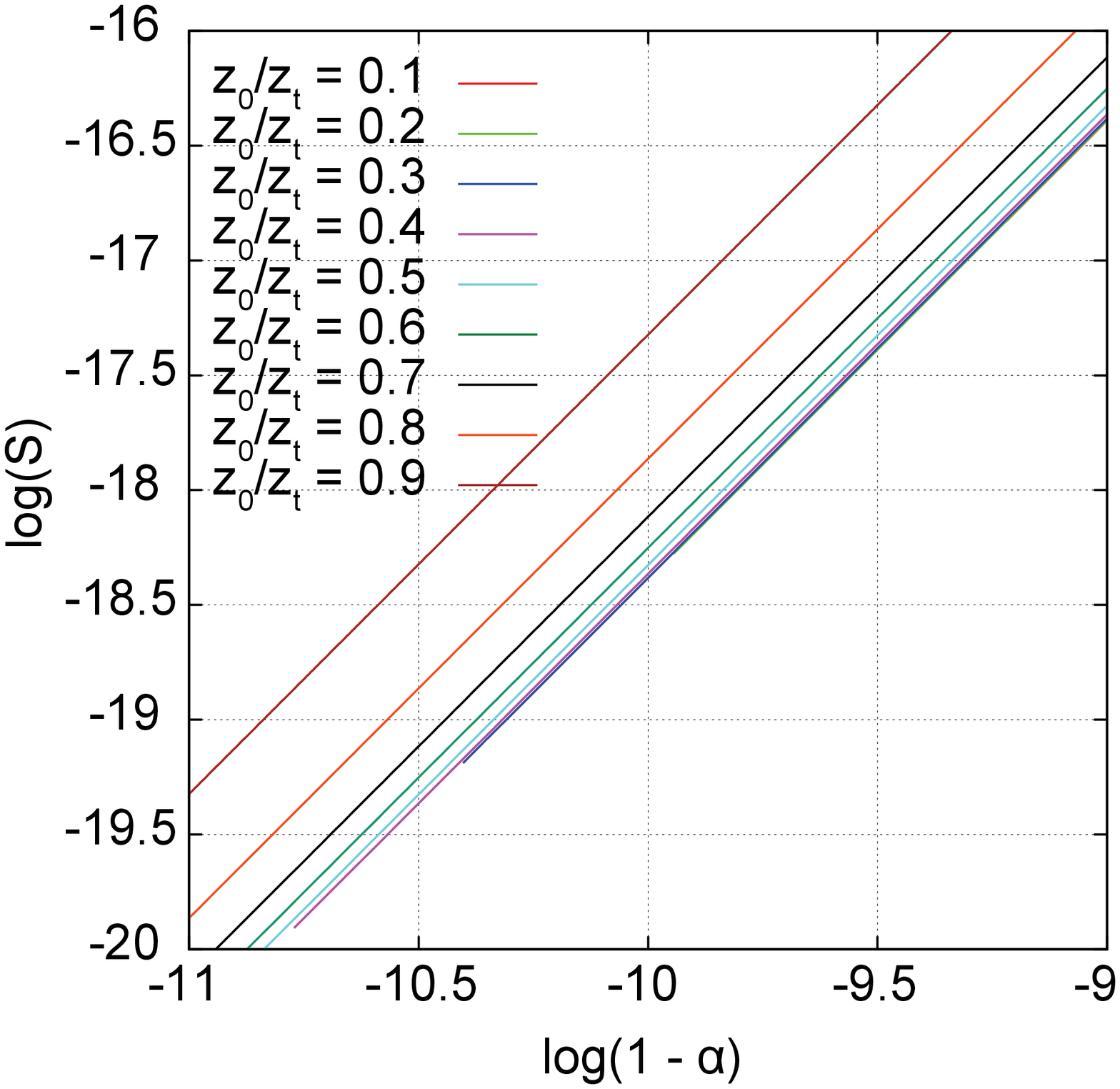} & 
\includegraphics[scale=.3]{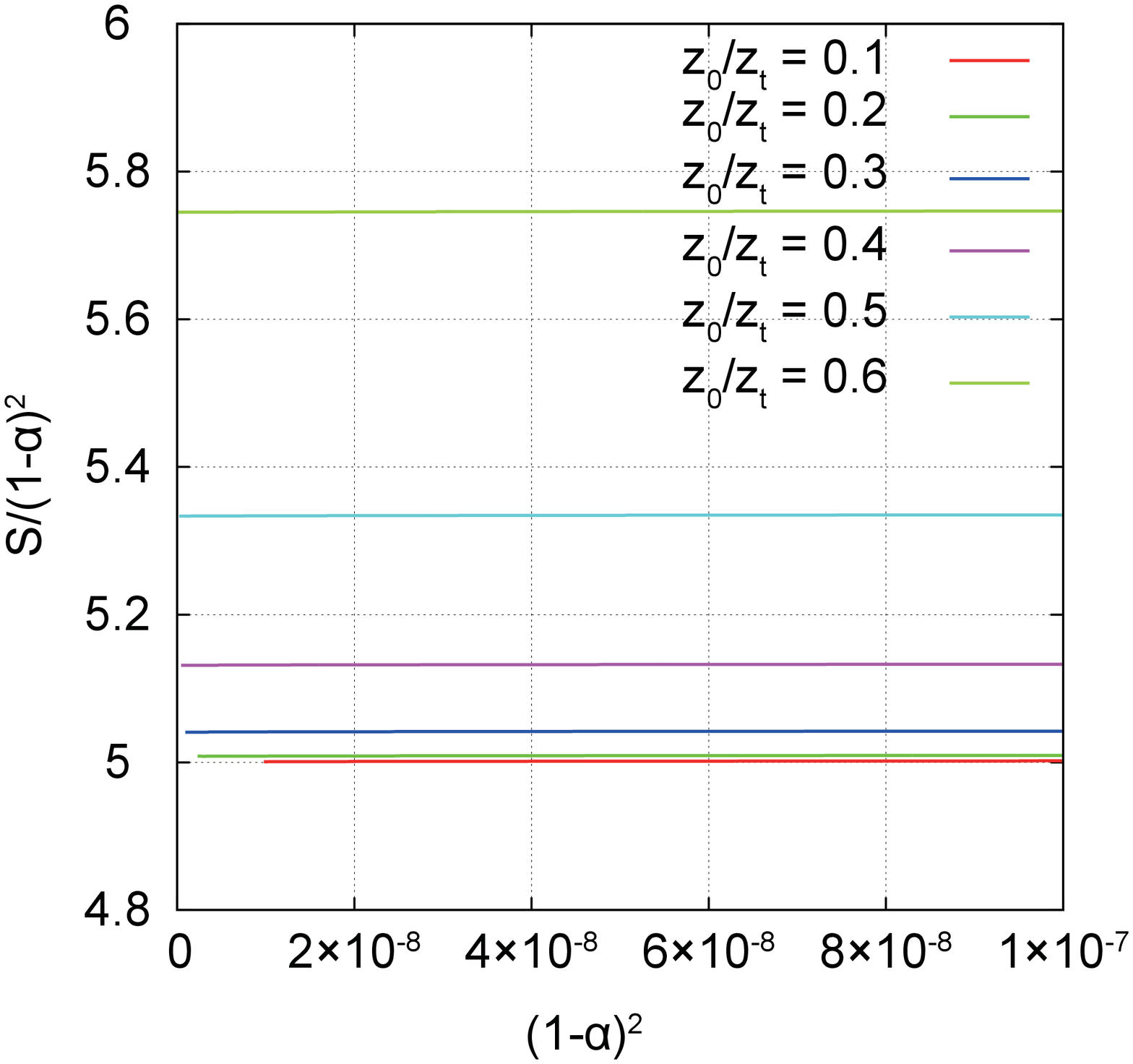} 
\vspace{-0.1cm}\\ 
\mbox{\footnotesize (a) the classical action } & \mbox{\footnotesize (b) ~~$B(\alpha_{\rm s})$}
\end{array}
\]
\caption{The behavior of the classical action near $E=E_{\rm c}$\,. 
The log-log plot in Fig.\ (a) indicates that $\gamma_{\rm c} =2$ universally. 
Figure (b) indicates that the coefficient approaches 5 (= $\frac{\sqrt{\lambda}}{2}$) 
as $\alpha_{\rm s} \to 0$ (Recall that $\lambda=100$).} 
\label{near-critical:fig}
\end{figure}

Let us next discuss the behavior near $E=E_{\rm c}$\,. 
The classical action should vanish as $\alpha \to 1$\,, and hence 
one can expect the following behavior: 
\begin{equation}
S = B(\alpha_{\rm s})(1-\alpha)^{\gamma_{\rm c}} + \cdots\,.
\end{equation}
Here $\gamma_{\rm c}$ is a positive constant to be determined and $B(\alpha_{\rm s})$ is an unknown function. 
The log-log plot of the classical action is shown in Fig.\,\ref{near-critical:fig} (a). 
The result indicates that the exponent $\gamma_{\rm c}$ is given by  
\[
\gamma_{\rm c} =2\,.
\]
Moreover, from Fig.\,\ref{near-critical:fig} (b), one can determine the asymptotic form of $B(\alpha_{\rm s})$ 
in the $\alpha_{\rm s} \to 0$ limit. As a result, the resulting classical action has been determined as 
\[
S = \frac{\sqrt{\lambda}}{2}(1-\alpha)^2 + \mathcal{O}\bigl((1-\alpha)^3\bigr)\,.
\]
Note that this expression completely agrees with the asymptotic form derived from (\ref{SZrate})  analytically. 
This agreement ensures the consistency of our numerical computations.

\section{Conclusion and Discussion}

We have reviewed the recent progress on a holographic description  
of the Schwinger effect in the Coulomb phase and the confining phase. 
The part of the Coulomb phase is mainly based on the seminal work by 
Semenoff and Zarembo \cite{SZ}. The part of the confining phase is 
a summery of a series of our works \cite{SY3,SY4,KSY}. 

\smallskip

First, we have introduced the world-line instanton method, 
which is a standard method to compute the Schwinger production rate. 
A remarkable point on this method 
is that the rate can be evaluated even at arbitrary coupling. 
Then a circular Wilson loop appears in the middle of the computation, 
in comparison to the weak-coupling analysis. 
This method is also applicable to the ${\cal N}\! =4$ SYM theory through the Higgs mechanism 
by assuming the heavy quark mass and the weak-field condition. 
Notably, in this case,  a circular 1/2 BPS Wilson loop appears in the formula of the production rate \eqref{n=4rate}. 
The VEV of the Wilson loop is estimated by using the AdS/CFT correspondence. 
However, there are unsatisfactory points: 
1) it seems unlikely that the Schwinger effect occur for very heavy quarks, 
2) disagreement of the critical electric fields computed from the production rate and 
the DBI action of a probe D3-brane. 

\smallskip

To resolve this problem, Semenoff and Zarembo improved the AdS/CFT set-up. 
The probe D3-brane is put at an intermediate position in the bulk AdS$_5$\,. 
Then the quark mass becomes arbitrary and the critical electric fields nicely agree. 
According to this improvement, the production rate is corrected with an additional 
term, which can be ignored under the weak-field condition. 
Note also that the Coulomb potential is modified \cite{SY2} as well. 
The critical electric field obtained from the potential analysis 
also agrees with the results from the production rate and the DBI action. 

\smallskip

The holographic method argued in the Coulomb phase has been generalized to confining gauge theories.
In this review we have focused upon an AdS$_5$ soliton background. 
The potential analysis has been done \cite{SY3} and 
it has been shown that there are two kinds of critical behaviors 
around (1) $E=E_{\rm s}$ and (2) $E=E_{\rm c}$\,. 
When $E \leq E_{\rm s}$\,, the Schwinger effect cannot occur due to the confining string tension. 
In the region with $E_{\rm s} < E < E_{\rm c}$\,, it is possible as a tunneling effect as usual. 
When $E \geq E_{\rm c}$\,, the vacuum is unstable catastrophically. 
Finally, the production rate has been evaluated numerically. 
Then we have introduced new critical exponents associated with the two critical behaviors. 
The values of the exponents are non-trivial results and 
might be regarded as a prediction in the real QCD via the universality argument, 
as in the case of the ratio of the shear viscosity $\eta$ to the entropy density $s$\,, $\eta/s =1/4\pi$ \cite{PSS}. 
Thus it is of importance to check the universality of them for various backgrounds \cite{Son1}. 
As a matter of course, it should be significant to reveal the mathematical foundation for the universality 
by employing the holographic approach discussed here as a compass. 





\smallskip 

We have concentrated on an AdS soliton background as a confining geometry. 
As other examples, one may consider $\mathcal{N}\! = 2$ supersymmetric QCD 
and the Sakai-Sugimoto model \cite{SS}. The Schwinger effect in the former case 
has been studied in a series of works \cite{Hashi,Hashi2,Hashi3}. 
As for the latter case, see the recent work \cite{Hashi4}. 
Another holographic Schwinger effect based on the bottom up approach 
is discussed by Dietrich \cite{Dietrich}. It would be significant to study 
the Schwinger effect in confining theories from various perspectives. 
It is also interesting to study a holographic description of
the Schwinger effect in de Sitter space \cite{dS}.
It is argued that no pair production occurs in a plane-wave background \cite{Sakaguchi}. 

\smallskip

We believe that the Schwinger effect in confining gauge theories 
would be a key ingredient in looking for new aspects of QCD in the presence 
of extremely strong external fields.

\section*{Acknowledgments}

We appreciate K.~Hashimoto and H.~Suganuma for useful comments and discussion. 
The work of YS is supported by a Grant-in-Aid for Japan Society for
the Promotion of Science (JSPS) Fellows No.\,26$\cdot$1300. 


\end{document}